\documentclass[11pt]{article}
\usepackage{graphicx} 
\usepackage[margin=1in]{geometry}
\usepackage[numbers, sort&compress]{natbib}
\usepackage{amsmath}
\usepackage{titlesec}
\usepackage{enumitem}
\usepackage[T1]{fontenc}
\usepackage{caption} 
\usepackage{titling}
\usepackage{chemformula}
\usepackage{setspace}
\usepackage{float}
\usepackage{gensymb}
\usepackage{amsfonts}

\usepackage{caption}

\usepackage{subcaption}
\usepackage[labelfont=bf]{caption}
\titlespacing*{\section}{0pt}{8.0pt}{0.5pt}
\titlespacing*{\subsection}{0pt}{8.0pt}{0.5pt}

\setlist[enumerate]{topsep=0pt,itemsep=-1ex,partopsep=1ex,parsep=1ex}
\setlist[itemize]{topsep=0pt,itemsep=-1ex,partopsep=1ex,parsep=1ex}

\title{\Large A transformer-based model for rapid microstructure inference from four-dimensional scanning transmission electron microscopy data\vspace{-0.3cm}}
\author{Kwanghwi Je, Ellis R. Kennedy, Sungin Kim, Yao Yang, Erik H. Thiede\textsuperscript{*}}
\date{}

\begin{document}

\maketitle

\renewcommand{\thefootnote}{\fnsymbol{footnote}}
\footnotetext{*Corresponding author: Erik H. Thiede (email: eht45@cornell.edu)}

\begin{abstract}
Properties of crystalline materials are closely linked to microstructure arising from the spatial arrangement, orientation, and phase of nanocrystals. Rapid characterization of crystalline microstructure can accelerate the identification of these links and the development of materials with desired properties. Here, we combine a machine learning framework with four-dimensional scanning transmission electron microscopy (4D-STEM) to enable fast inference of crystalline microstructure over large fields of view. The framework employs a transformer-based architecture to predict crystallographic orientations and phases from 4D-STEM diffraction patterns, yielding spatially resolved maps of microstructural features at the nanoscale. With this framework, crystallographic orientations are inferred up to two orders of magnitude faster than widely used correlative template-matching approaches. This capability enables high-throughput characterization of complex crystalline materials and facilitates the establishment of structure–property relationships central to materials design and optimization.

\end{abstract}


\section*{Introduction}

Properties of crystalline materials are governed by their microstructures, which arise from the spatial arrangement of nanometer-scale crystals together with their crystallographic phases and orientations. Multiple microstructures have demonstrated functionalities in mechanical \cite{cheng2018extra,zhou2018enhanced,fang2011revealing,bauer2014high}, acoustics \cite{kim2017phonon,weaver1990diffusivity}, and electromagnetics \cite{bishara2021understanding,xu2020charge,geng2024grain} applications. To establish structure-property relationships and enable the rational design of crystalline materials, robust characterization of the crystalline microstructure at fine length scales is essential. Four-dimensional scanning transmission electron microscopy (4D-STEM) has proven highly versatile for meeting these characterization needs \cite{liu20244d}. In 4D-STEM, a focused electron probe is scanned across a two-dimensional area of a specimen, and a diffraction pattern is recorded at each scan position with nanometer-scale spatial resolution \cite{ophus2019four}. For crystalline materials, each recorded pattern contains Bragg disks that encode local crystallographic orientations and phases \cite{folastre2024improved}. 

Determining the encoded crystallographic orientations and phases across the scan grid provides spatially resolved maps of oriented nanocrystals at high spatial resolution, which enable detailed characterization of crystalline microstructure. However, achieving fast and robust determination of these two structural attributes remains challenging because of the large size and complexity of 4D-STEM datasets. The diffraction patterns contain rich but complex features that are difficult to analyze reliably at scale, and human-driven inspection becomes impractical for large 4D-STEM data volumes. These limitations motivate the development of automated approaches that can reliably map structural attributes across extensive 4D-STEM measurements.

One powerful approach for automating the mapping is correlative template matching \cite{rauch2010automated,rauch2014automated,ophus2022automated,cautaerts2022free}. In this framework, each 4D-STEM diffraction pattern is compared with a large library of simulated templates using correlation scores. The template with the highest score is selected, and its associated crystallographic orientation or phase is assigned to the pattern. Repeating this procedure across all scan positions enables automatic mapping of the structural attributes across the full 4D-STEM dataset. Correlative template matching has become widely used in the microscopy community and has been extended to infer additional material features \cite{morawiec2007algorithm}. Despite its widespread use, the approach scales poorly because the computational cost of performing these pairwise comparisons increases rapidly with the size of both the 4D-STEM dataset and the template library. As a result, there is a strong need for methods that mitigate these scaling limitations while still accurately predicting the structural attributes across the full dataset.

In this work, we introduce a transformer-based model for rapid inference of crystalline microstructure from 4D-STEM data. Our approach directly maps individual diffraction patterns to their corresponding structural attributes by treating Bragg disks as discrete tokens and exploiting relationships among Bragg disks. By replacing the exhaustive pairwise comparison with a fixed sequence of learned transformations, the model avoids the substantial computational cost associated with correlative template matching and is computationally scalable. As a first demonstration of the model’s capabilities, we apply it to orientation mapping of 4D-STEM diffraction patterns and evaluate its performance relative to correlative template matching. This design reduces the computation time required for orientation mapping by up to two orders of magnitude compared with a widely used correlative template matching approach. For simulated 4D-STEM diffraction patterns, the predicted orientations differ from the true orientations by a mean geodesic distance of $0.013$, corresponding to an average angular misalignment of $0.013$ radians. This level of angular agreement indicates a high accuracy of the orientation predictions.

To evaluate the performance of the model on a demanding experimental dataset, we performed orientation prediction on noisy 4D-STEM data collected from dendritic copper ($Cu$) crystals grown in liquid under an applied electric field. The diffraction patterns from these complex architectures are highly noisy and contain only a small number of Bragg disks, making orientation mapping particularly challenging. Despite the noise and limited diffraction signals, our model provides orientation predictions with moderate accuracy and captures crystalline domain structure, demonstrating that the model remains functional on noisy experimental 4D-STEM measurements.

In addition to predicting orientations, our model can be extended to further predict crystal phases, yielding spatially resolved maps of crystalline microstructure. We demonstrate that the model can jointly predict both crystal orientations and phases of individual $Cu$ and cuprous oxide ($Cu_{2}O$) crystal grains from a synthetic 4D-STEM dataset. Our choice of application is motivated by the recent results showing that the activity of $Cu$-based electrocatalysts is strongly influenced by the spatial distribution, coexistence, and grain structure of $Cu$ and $Cu_{2}O$ phases\cite{aran2020role,cheng2025direct}. While we focus here on catalytic nanoparticles as a representative case, the methodology is broadly applicable to materials systems characterized by coexisting phases and multiple orientations. Efficient extraction of these structural attributes will advance high-throughput microstructure analysis of 4D-STEM data and support the identification of structure–property relationships central to the development of complex functional crystalline materials.


\section*{Results}

\subsection*{Model framework and orientation prediction}

Our transformer-based model infers structural attributes from Bragg disks in a diffraction pattern. The model contains two neural network components: an encoder-only transformer followed by a multilayer perceptron (MLP) head. The first network component is a transformer encoder (Fig. 1a, top right) that operates on a set of Bragg disk embeddings (Fig. 1a, left), which are vector representations of individual disks. In a diffraction pattern, the structural attributes are reflected in the positions and intensities of Bragg disks and in the relationships among them. An analogous situation occurs in natural language, where the meaning of a sentence depends on both the information carried by individual word tokens and their contextual relationships. Transformers are designed to capture both individual token information and contextual relationships by using the attention mechanism \cite{vaswani2017attention}. In our framework, we adopt this formulation by representing each Bragg disk as a token and using a transformer encoder to predict these attributes from the combined token information. For each disk, we construct embeddings from its two-dimensional position, expressed in polar coordinates $k_{r}$ (radial distance) and $k_{\theta}$ (polar angle), and its intensity $I$. The three embeddings are summed to form the embedding for each disk. The transformer encoder operates on the disk embeddings to produce contextualized representations that capture relationships among disks within each diffraction pattern (see \textit{Supplementary Text S1} for a more detailed discussion of contextual learning for our model). These representations are then combined by mean pooling to produce a single latent vector that represents the diffraction pattern. 

The second network component of the framework is a multilayer perceptron (MLP) head that maps the latent vector to the structural attributes of interest (Fig. 1a, bottom right). As a first demonstration of the framework, we focus on predicting crystal orientation. For this task, the MLP head maps a latent vector to a crystal orientation represented as a proper rotation matrix in $\mathrm{SO}(3)$ (Fig. 1b right).

\begin{figure}[H]
\begin{center}
\includegraphics[width=0.995\textwidth]{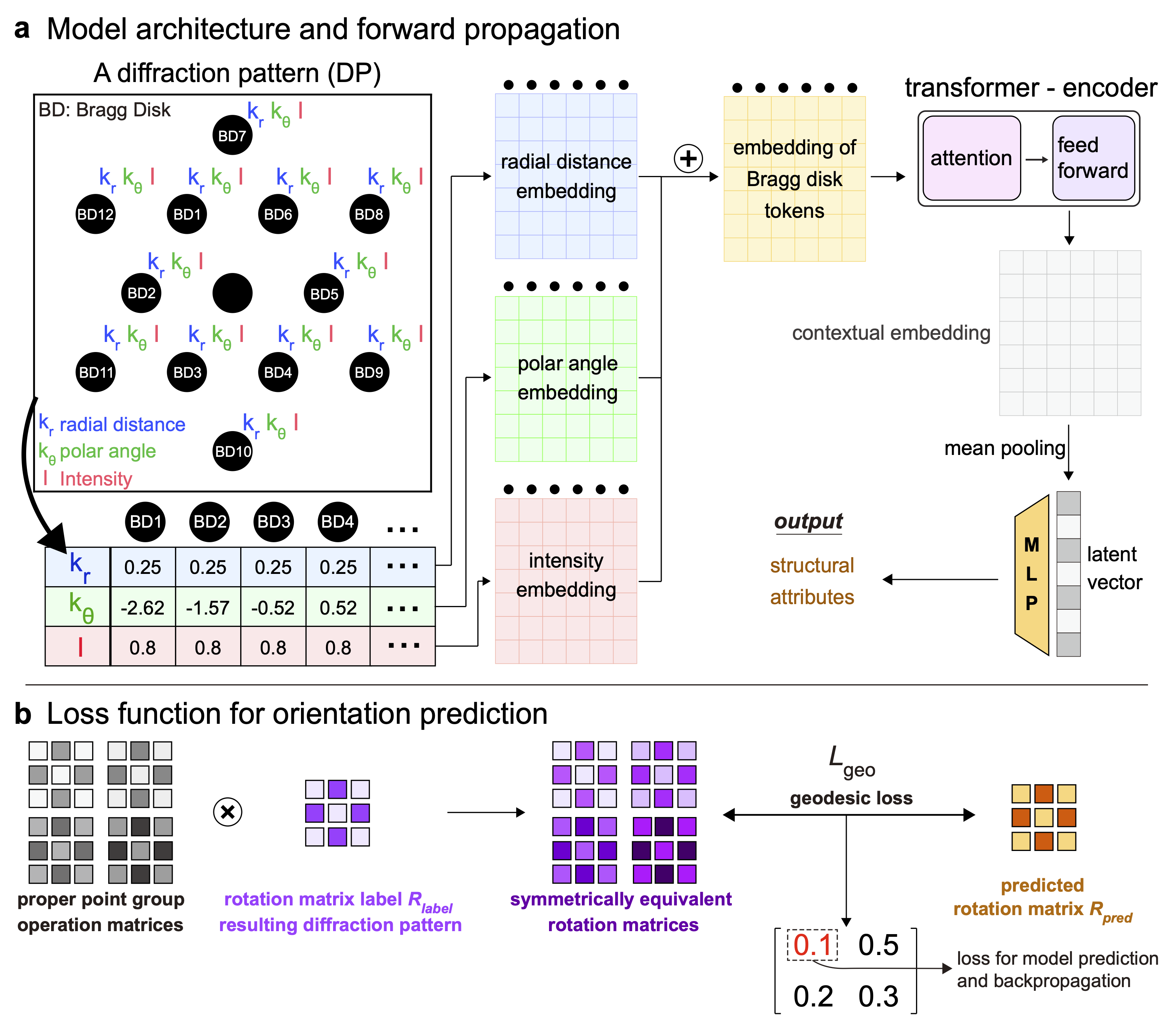}
\vspace*{-3mm}
\caption{Schematic illustration of model and prediction workflow. \textbf{a} Each Bragg disk in the diffraction pattern is treated as a token whose radial distance $k_{r}$, polar angle $k_{\theta}$, and intensity $I$ are embedded and summed to yield a token representation. A transformer encoder processes the set of tokens to produce contextualized embeddings, which are then combined into a latent vector using mean pooling. A multilayer perceptron (MLP) head maps the latent vector to target structural attributes. \textbf{b} Symmetry-aware geodesic loss $L_{\mathrm{geo}}$ used to train the model for orientation prediction. The MLP head maps the latent vector to a rotation matrix $R_{\mathrm{pred}}$ in $\mathrm{SO}(3)$, representing the crystal orientation. Each diffraction pattern in the training set is simulated from an orientation label $R_{\mathrm{label}}$, chosen as a representative of its symmetry-equivalent orientation class. During training, all symmetry-equivalent variants of the label are generated using proper point group operators of crystals. The predicted orientation is compared with these variants using the geodesic distance on $\mathrm{SO}(3)$, which quantifies the angular misorientation between rotations. The minimum distance defines the loss.}
\label{fig:FIG_1}
\end{center}
\end{figure}


In our orientation prediction framework, each diffraction pattern in the training set is paired with an orientation label represented as a rotation matrix $R_{\mathrm{label}}$. Because crystals possess proper point group symmetries, multiple orientations related by these symmetry operations can produce indistinguishable diffraction patterns. To account for this symmetry during training, we use a symmetry-aware geodesic loss $L_{\mathrm{geo}}$ (Fig. 1b). After the model prediction, geodesic distances on the rotation manifold $\mathrm{SO}(3)$ are measured between the predicted orientation and each symmetry-equivalent variant of the orientation label obtained by applying the crystal’s proper point group symmetry operations. The geodesic distance corresponds to the smallest rotation angle required to align the two orientations. The symmetry-aware geodesic loss $L_{\mathrm{geo}}$ for a given diffraction pattern is defined as the minimum of these distances.

To validate the model's ability to predict crystal orientations, we trained the model using synthetic diffraction patterns of face-centered-cubic (fcc) $\textit{Cu}$ crystals (space group $Fm\bar{3}m$, Fig. S1a).
We generated a training and validation set of $\textit{Cu}$ diffraction patterns from a set of symmetrically unique crystal orientations and a range of specimen thicknesses (Fig. 2a and Supplementary Text S2) using dynamical Bloch-wave simulations implemented in the py4DSTEM software (version 0.14.08) \cite{savitzky2021py4dstem}. The Bragg disks in the simulated pattern serve as the model input, and the orientation used in the simulation serves as the orientation label. We introduced data augmentations during training, including removing weak reflections and perturbing disk positions and intensities (Fig. S2 and \textit{Supplementary Text S3}), to make the model robust to Bragg disk feature variations present in experimental patterns.

To obtain an unbiased estimate of the model’s performance on unseen 4D-STEM diffraction data, we generated an independent synthetic test set from randomly sampled crystal orientations (Fig. 2b) and specimen thicknesses (Fig. S3). This approach yields a test set that is statistically independent of the training data and supports a robust evaluation of orientation prediction accuracy.

\begin{figure}[H]
\begin{center}
\includegraphics[width=0.995\textwidth]{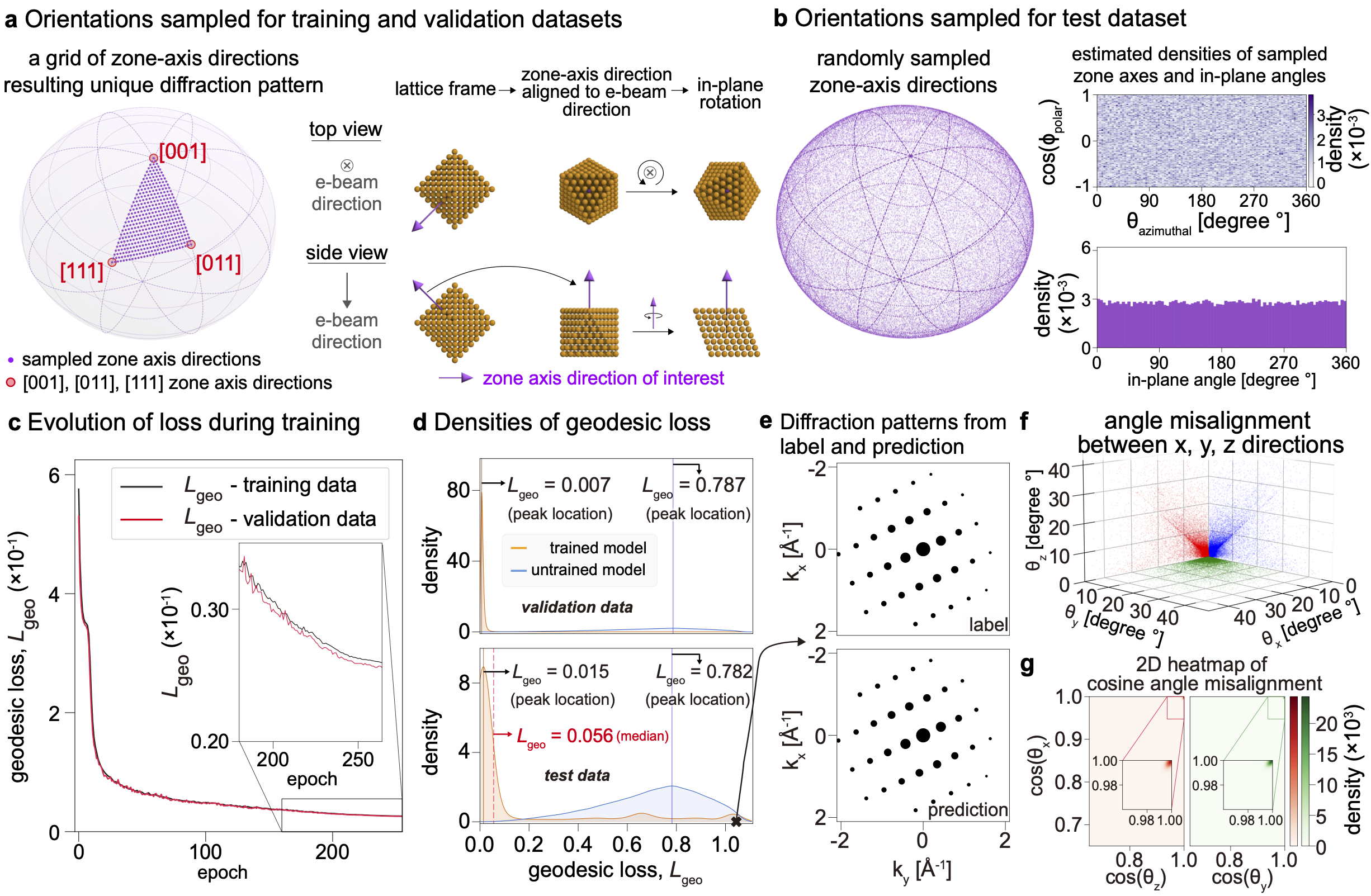}
\vspace*{-3mm}
\caption{\textbf{a} Sampling of orientation labels for generating the training and validation sets. These labels are sampled in two steps. A grid of symmetrically unique zone-axis directions is sampled for the selected face-centered-cubic (fcc) copper ($Cu$) crystal (left). These directions lie on the surface of the unit sphere bounded by the [001], [011], and [111] directions. For each zone axis, the crystal lattice is rotated to align the zone axis with the incident electron beam. An additional in-plane rotation about the zone axis is then applied to sample the full range of crystal orientations (right). \textbf{b} Sampling of orientation labels for generating the test set. Zone-axis directions are randomly sampled on the surface of the unit sphere (left), yielding an approximately uniform distribution in $\cos(\phi_{\mathrm{polar}})$ and $\theta_{\mathrm{azimuthal}}$ (top right). $\phi_{\mathrm{polar}}$ and $\theta_{\mathrm{azimuthal}}$ denote the polar and azimuthal angles in spherical coordinates. Uniformly sampled in-plane rotations (bottom right) are then applied about each zone axis. \textbf{c} Evolution of geodesic loss $L_{\mathrm{geo}}$ for the training and validation sets during training. \textbf{d} Estimated densities of $L_{\mathrm{geo}}$ for the validation set (top) and test set (bottom), shown for the trained model (brown) and an untrained model (blue). \textbf{e} Diffraction patterns simulated from a reference orientation label and from the corresponding predicted orientation for a data point marked in (d). The two patterns exhibit high visual similarity despite a large $L_{\mathrm{geo}}$ value. \textbf{f} Distribution of angular misalignment between symmetry-reduced crystal axes for the orientation label and predicted orientation, $\theta_{x}$, $\theta_{y}$, and $\theta_{z}$. \textbf{g} Estimated density of the cosine of the angular misalignment in (f).}
\label{fig:FIG_2}
\end{center}
\end{figure}


During model training, the symmetry-aware geodesic loss $L_{\mathrm{geo}}$ for both training and validation data gradually decreased and reached a small value below $0.03$ radians (Fig. 2c). After training, the density of $L_{\mathrm{geo}}$ for the validation dataset shows a sharp peak at $0.007$ radians (Fig. 2d, top), and the mean geodesic loss across the dataset is $0.013$ radians, indicating that the predicted orientation is close to one of the full set of symmetry-equivalent orientations on the $\mathrm{SO}(3)$ manifold. For the test set, the average $L_{\mathrm{geo}}$ was $0.159$ radians, with the peak of the $L_{\mathrm{geo}}$ distribution at $0.015$ radians (Fig. 2d, bottom). While the model predicts orientations moderately well for the majority of unseen test data, a considerable number of predictions yield a high geodesic loss. Part of this discrepancy arises because the orientations used for the training and validation data were sampled differently from those used for the test data. Beyond those related by proper point group symmetry, multiple orientations can result in diffraction patterns that are identical or very similar; such orientational ambiguity often arises from projection-induced rotational symmetries or crystallographic index permutations associated with the two-dimensional projection of the three-dimensional reciprocal lattice. To prevent training conflicts arising from this ambiguity, we retained only a single representative orientation as the label for each diffraction pattern in the training and validation data (see \textit{Methods} and \textit{Supplementary Text S2}).

In contrast, the orientations in the test set were sampled completely randomly across the $\mathrm{SO}(3)$ space. This random sampling included orientations that were intentionally excluded from the labels used for the training and validation datasets but still result in diffraction patterns nearly identical to those in these datasets. Consequently, even when the model accurately predicts an orientation ($R_{\mathrm{pred}}$) for a given input diffraction pattern, that orientation may not be symmetry-equivalent to the specific, randomly sampled test label. This disparity creates a non-overlapping region between the model's predictions and the test set labels, sometimes resulting in high geodesic loss. An example of such a case is shown in Fig. 2e, where the model prediction yields an $L_{\mathrm{geo}}$ of $1.044$ radians despite the two simulated patterns from the test label and model prediction showing high visual similarity. To evaluate the performance of model prediction while accounting for the disparity, we applied symmetry-reduction to the test set label orientations and predicted orientations, and compared the axis-wise angular difference between the symmetry-reduced orientations. The angular difference between the symmetry-reduced principal crystal axes (i.e., the x-, y-, and z-directions) from the two orientation sets was highly populated below 1.5 degrees (Fig. 2f,g), which indicates that the principal crystal directions in the two orientation sets are highly aligned (see also Fig. S4). Overall, these results demonstrate that our model predicts orientation accurately on simulated diffraction data, and the observed high $L_{\mathrm{geo}}$ values in the test set can in some cases be ascribed to orientation ambiguities induced by the two-dimensional projection.


\subsection*{Benchmarking computation time for 4D-STEM orientation mapping}

The model performs the orientation mapping task substantially faster than correlative template matching. We benchmarked the computation time of both approaches on synthetic 4D-STEM datasets on scan grids ranging from $8\times8$ to $1024\times1024$ positions (Fig. 3a). For a fair comparison, both methods used identical inputs and generated outputs in the same format (Fig. 3b,c). The input was preprocessed 4D-STEM data in which each scan position is associated with a set of simulated Bragg disks representing a synthetic diffraction pattern. The output is an orientation map where each scan position is assigned a $\mathrm{SO}(3)$ crystal orientation.

Orientation mapping with model predictions was evaluated in both CPU-only and GPU modes. For template matching, we used the Automated Crystal Orientation Mapping (ACOM) method \cite{ophus2019four} implemented in the py4DSTEM software, which was executed on CPU cores in the present study. We used default parameters for the ACOM template matching method. Wall time measurements for all dataset sizes are shown in Fig. 3d. Compared with correlative template matching, the model achieved substantial reductions in computation time for scan grids larger than $64\times64$ positions. For example, for a dataset with a $512\times512$ grid of scan positions, the model required 358 s in CPU-only mode and 53 s in GPU mode to complete the mapping, while correlative template matching required 5173 s on the CPU. This corresponds to an acceleration of roughly 14 times on CPU-only and 98 times on GPU-accelerated inference when compared with template matching.

With GPU acceleration and for scan grids larger than $64\times64$ positions, the time required for model prediction became shorter than that for loading the 4D-STEM data into the computation environment during orientation mapping (Fig. 3e). These results indicate that model inference is not the bottleneck in the orientation mapping of large 4D-STEM datasets. The rapid inference of the model can support orientation mapping across many 4D-STEM measurements and thus can facilitate high-throughput microstructure analysis. The computing platforms and benchmark settings are summarized in the \textit{Methods} section.
\newline

\begin{figure}[H]
\begin{center}
\includegraphics[width=0.99\textwidth]{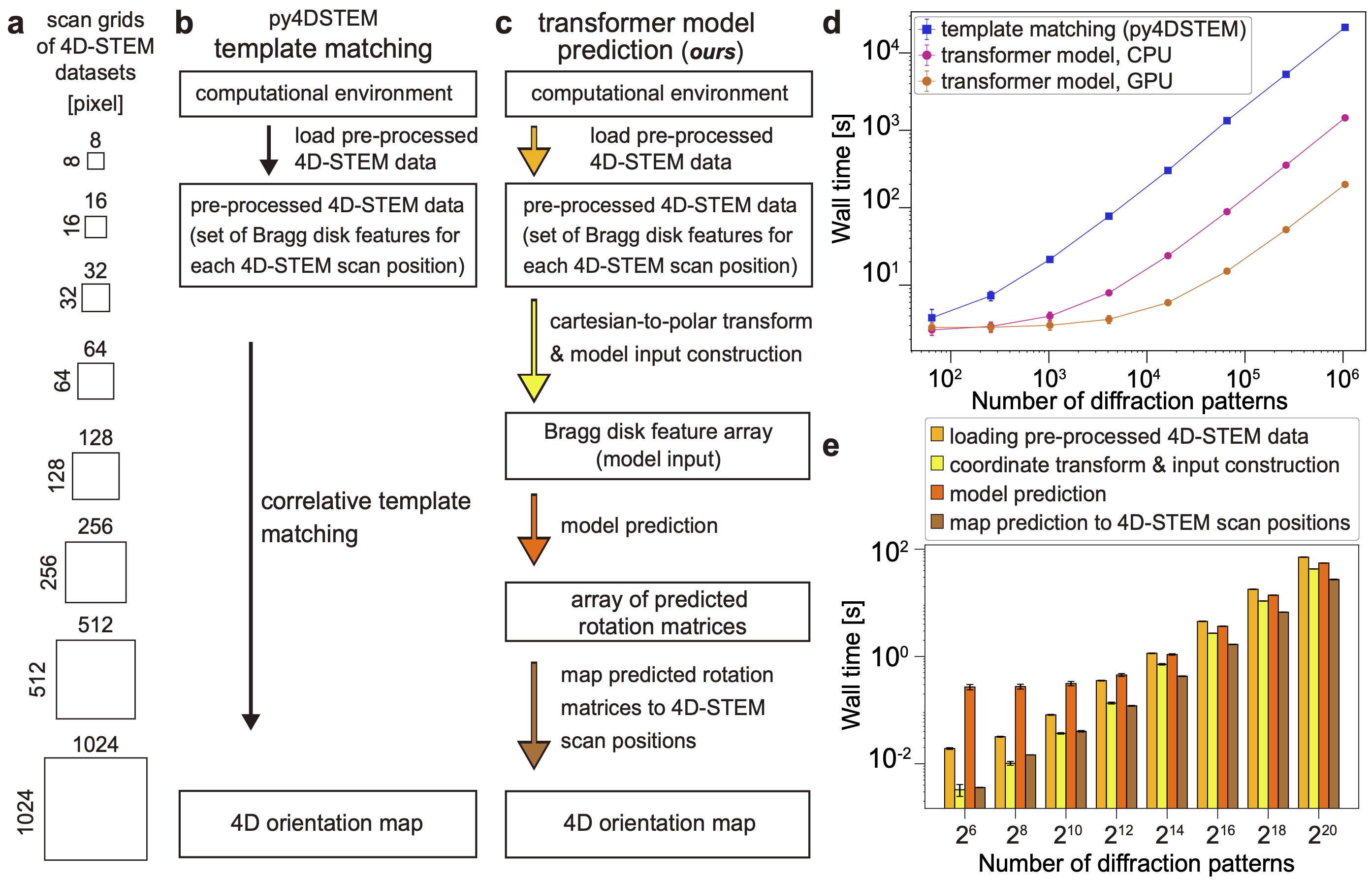}
\vspace*{-3mm}
\caption{\textbf{a} Four-dimensional scanning transmission electron microscopy (4D-STEM) datasets with different scan grid sizes used in this study to benchmark computation times for orientation mapping. \textbf{b} Orientation mapping of 4D-STEM data using correlative template matching implemented in py4DSTEM. \textbf{c} Orientation mapping of 4D-STEM data using our model, which involves four sequential steps: loading pre-processed 4D-STEM data (Bragg disk feature map across a four-dimensional grid of scan positions), transforming Bragg disk positional coordinates and constructing the model input, performing model inference, and assembling the final orientation map. \textbf{d} For each 4D-STEM dataset and method, we measured the elapsed time for orientation mapping using 10 independent runs and report the corresponding mean and standard deviation. \textbf{e} Wall-clock execution times for the four processes in (c) when using the GPU. Elapsed time for each step was measured independently, and statistics were obtained over 10 repeated runs as in (d).}
\label{fig:FIG_3}
\end{center}
\end{figure}


\subsection*{Predicting orientations from an experimental 4D-STEM dataset of $Cu$ crystals}

After confirming reliable orientation predictions on simulated diffraction patterns, we applied our model to noisy experimental 4D-STEM data of dendritic \textit{Cu} crystals with fcc symmetry to assess its performance for experimental measurements (Fig. 4a). The $\textit{Cu}$ dendrites were grown in liquid under an applied electric field \cite{kim2025operando}. Diffraction patterns from these complex structures exhibit a low signal-to-noise ratio and contain only a small number of Bragg disks, which makes reliable orientation prediction challenging. This dataset provides a stringent test of our model and allows us to evaluate its ability to predict orientations in the presence of substantial noise. As a reference, the ACOM template matching was used for comparison.

We obtained orientation maps for the 4D-STEM data from both the template matching and our model predictions (Fig. 4b,c). In each map, two symmetry-reduced crystal lattice directions are derived from the predicted orientation at each scan position and presented through color coding (the inset of the right panel of Fig. 4c). The directions correspond to one in-plane direction and one out-of-plane direction. We note that the symmetry-reduction was performed using the full crystallographic point group as implemented in py4DSTEM.

For some crystalline domains, the orientations predicted by the two methods agree well, which is evident from the similar color assignments across the corresponding scan positions. However, in other regions, the two maps display noticeably different color representations and accordingly different orientation predictions. One source of the difference is the substantial noise in many diffraction patterns, which can contribute to incorrect orientation assignments in one or both methods. This issue is not solely due to differences in the orientation predictions from the two methods, because a similar issue appears within each map. In each individual map, the color representation frequently changes from one scan position to the next. Such abrupt changes imply large variations in the predicted orientations at neighboring scan positions. Because crystals at neighboring scan positions have similar orientations, the color assignments are, in general, consistent or vary smoothly across neighboring positions. The persistent lack of coherence in both maps indicates that the underlying experimental data are extremely noisy, and that both approaches sometimes struggle to produce accurate orientation predictions.

\begin{figure}[H]
\begin{center}
\includegraphics[width=0.99\textwidth]{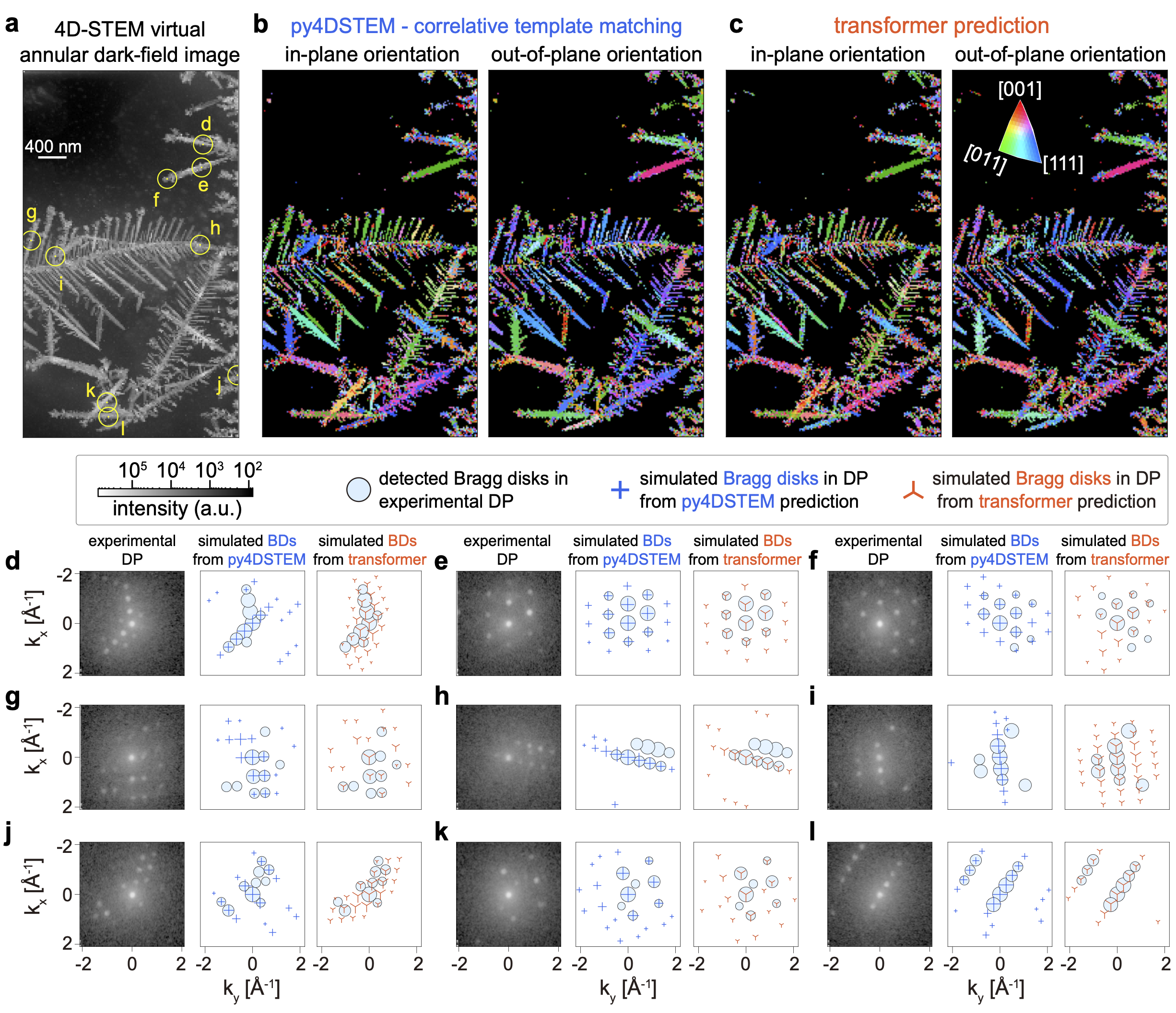}
\vspace*{-3mm}
\caption{\textbf{a} A virtual 4D-STEM annular dark-field image of fcc $Cu$ dendritic crystals grown in liquid under an electric field. \textbf{b,c} Orientation maps of 4D-STEM data in (a) obtained from py4DSTEM template matching (b) and from our model (c). Left and right panels are maps of in-plane orientation and out-of-plane orientation, corresponding to symmetry-reduced x- and z-axis directions, respectively. The orientation color code is shown in the right panel of (c). Black regions indicate scan positions for which no orientation is assigned. \textbf{d--l} Comparison of predictions from py4DSTEM template matching and from the model for selected diffraction patterns marked by yellow circles in (a). For each example, the experimental diffraction pattern (left), the simulated pattern from py4DSTEM prediction (middle), and the simulated pattern from the model prediction (right) are shown. To illustrate the correspondence between simulated and experimental Bragg disks, the experimental Bragg disks (blue) are overlaid on the simulated diffraction patterns. The selected diffraction patterns in (a) are randomly drawn from those containing more than six Bragg disks.}
\label{fig:FIG_4}
\end{center}
\end{figure}


Despite the challenges in obtaining reliable orientation predictions and the absence of ground-truth labels for the 4D-STEM data, we can still assess model performance by simulating diffraction patterns from the predicted rotations and comparing with their experimental counterparts. Fig. 4d-l shows randomly sampled diffraction patterns (left), corresponding Bragg disks simulated from py4DSTEM template matching predictions (middle), and those simulated from model predictions (right). The more similar the experimental Bragg disks and the corresponding simulated Bragg disks are, the more likely the predictions are to be acceptable. 

To provide a quantitative comparison and evaluate the prediction performance, we measured the sparse correlation score \cite{ophus2019four} and cross-correlation score $Q$ \cite{rauch2014automated,jeong2021automated,cautaerts2022free,folastre2024improved} between the experimental and simulated Bragg disks; a graphical description is shown in Fig. S5. The sparse correlation score quantifies the similarity between compact polar-coordinate tables that encode the positions and intensities of the Bragg disks. It is the correlation metric used to find the best matching simulated template in the py4DSTEM template matching. The cross-correlation score $Q$ is computed as the sum of the products of the Bragg disk intensities in a simulated template and the corresponding intensities in the experimental diffraction pattern. Fig. 5a shows the estimated density of the two correlation scores for the model predictions (red) and for the py4DSTEM template matching predictions (blue) obtained from the experimental 4D-STEM data in Fig. 4a. For many diffraction patterns, both methods yield high $Q$ scores, indicating good agreement between the simulated and experimental Bragg disks and reasonable orientation estimates from each approach. Across diffraction patterns, $Q$ scores from the model predictions and template matching exhibit an approximately linear relationship (Fig. S6). While this trend holds broadly, in the low-$Q$ regime, model predictions yield lower correlation scores than template matching for some diffraction patterns. For the sparse correlation score, the values obtained from the model predictions are markedly lower than those from template matching for the majority of diffraction patterns. In the following paragraphs, we discuss factors that can give rise to the relatively low sparse correlation scores.

One factor that contributes to the lower correlation scores is that the model sometimes returns relatively less accurate predictions than template matching; in Figure 5b we have depicted six patterns specifically chosen to show poorer performance by our model. 
These examples suggest that our model often gives worse results when the number of detected Bragg peaks is very limited. For many of these cases, the simulated patterns from both the transformer and template matching do not visually match the experimentally observed pattern, suggesting that neither prediction is correct. However, the two approaches fail in different ways: whereas the transformer gives an incorrect prediction, template matching identifies an orientation whose corresponding pattern exhibits at least partial agreement with the detected Bragg disks.

Inaccurate model predictions are not the only factor that can produce the relatively low correlation score. Another factor is the way the sparse correlation score is calculated. Even when all experimental disks are assigned to the simulated Bragg disks from the model predictions, small offsets between the predicted disks and the true disks can result in relatively low sparse correlation scores (Fig. 5c right panels). In contrast, template matching can still return relatively high correlation scores for patterns where only a subset of the experimental Bragg disks align well with a simulated template (Fig. 5c middle panels). These higher correlation scores do not necessarily indicate that template matching always provides more accurate orientation estimates than the model. Rather, they reflect that template matching can sometimes yield moderate to high correlation scores even when its orientation predictions are relatively inaccurate. Overall, the model provides reasonable orientation estimates for some noisy experimental diffraction patterns (Fig. S7), although template matching performs better on average for this dataset.

\begin{figure}[H]
\begin{center}
\includegraphics[width=0.97\textwidth]{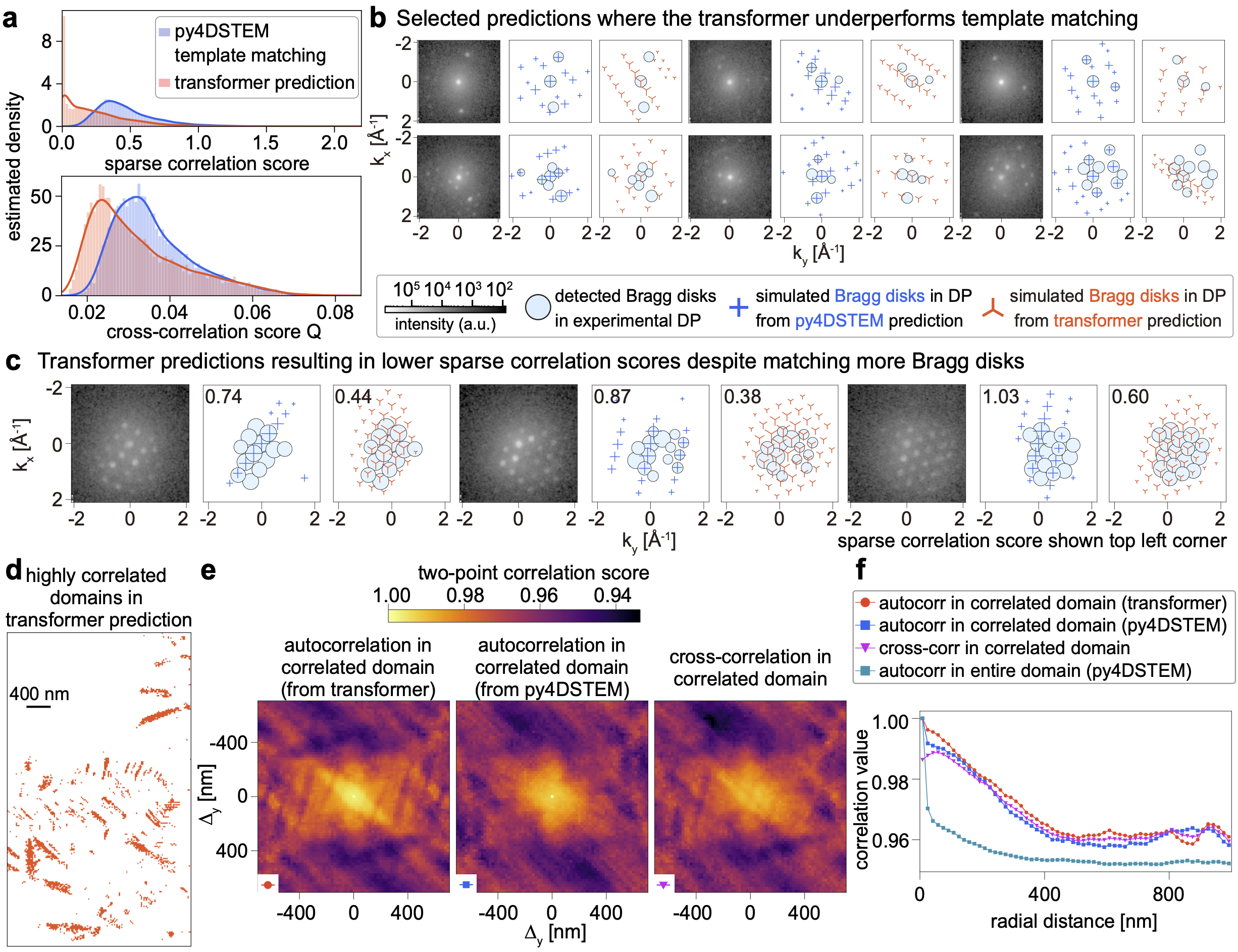}
\vspace*{-3mm}
\caption{\textbf{a} Estimated density of the sparse correlation score (top) and the cross-correlation score $Q$ (bottom) from model predictions (red) and py4DSTEM template matching (blue). \textbf{b} Selected examples where transformer predictions underperform template matching. Experimental diffraction patterns (left), simulated patterns from py4DSTEM template matching (middle), and simulated patterns from the model predictions (right). The examples are selected from cases in which model predictions yield lower sparse correlation scores than those obtained from template matching. \textbf{c} Examples illustrating why model predictions yield lower sparse correlation scores than template matching despite a larger number of matched Bragg disks. Bragg disks derived from model predictions show less precise positional agreement with the experimental disks, whereas those derived from template matching exhibit closer positional or intensity alignment for a subset of the experimental disks. Sparse correlation scores are shown in the top left of each panel. \textbf{d} Highly correlated domain identified from the model predictions. The domain corresponds to scan positions whose predicted zone axes show strong angular alignment with those of their neighboring scan positions. \textbf{e} Two-point correlation of predicted zone axes for positions in the domain in (d). The panels display the autocorrelation of the model predictions (left), the autocorrelation of the py4DSTEM predictions (middle), and the cross-correlation between the model and py4DSTEM predictions (right). \textbf{f} Azimuthal average of the two-dimensional correlations in (e). As a reference, we also show the autocorrelation of py4DSTEM zone-axis predictions over the entire scan region where orientations are assigned.}
\label{fig:FIG_5}
\end{center}
\end{figure}


\begin{figure}[H]
\begin{center}
\includegraphics[width=0.99\textwidth]{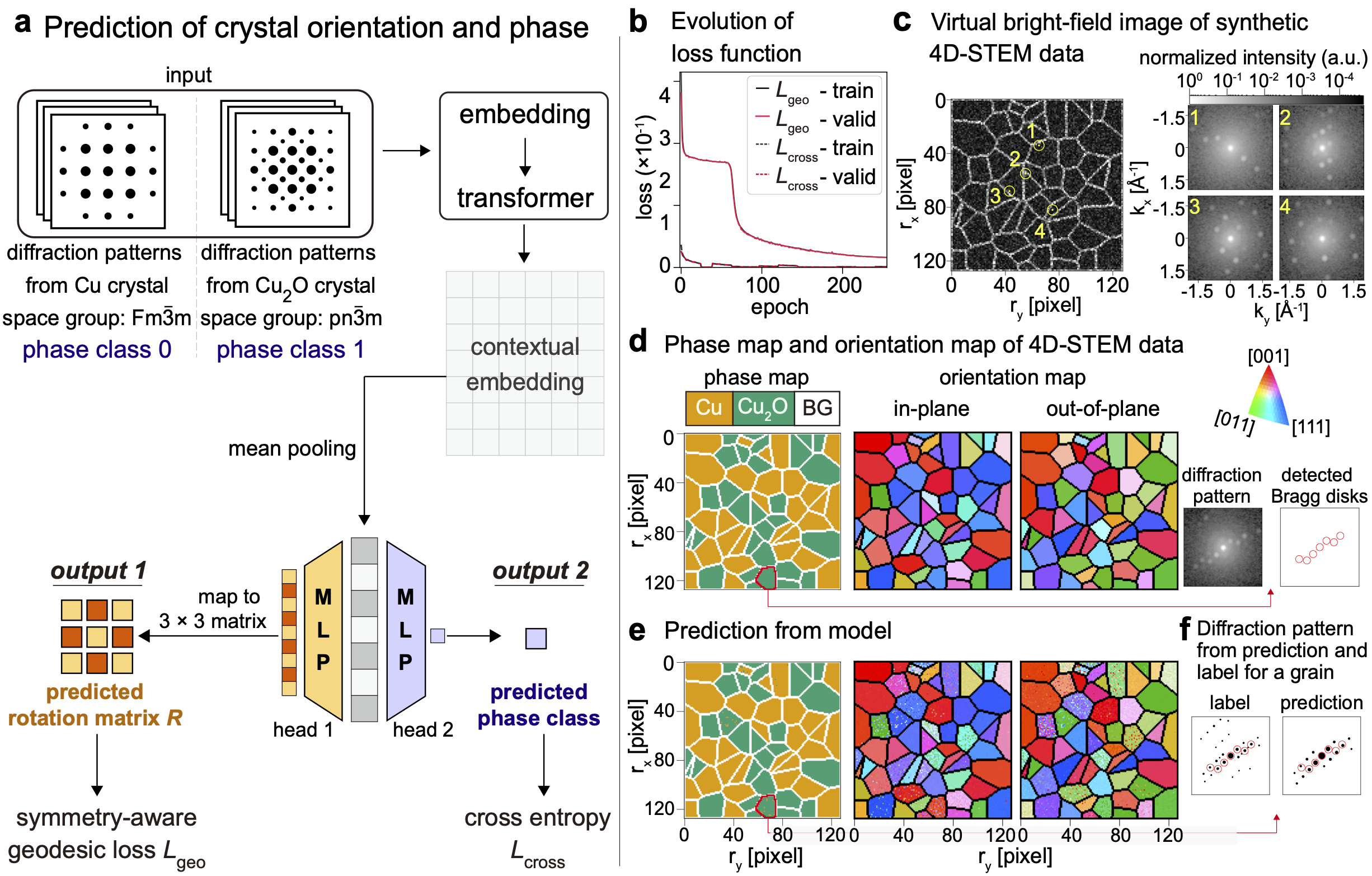}
\vspace*{-3mm}
\caption{\textbf{a} Schematic illustration of the model used for joint crystal phase and orientation prediction. Training and validation data consist of simulated diffraction patterns from fcc $Cu$ and cuprous oxide ($Cu_{2}O$) crystals over a range of orientations. Two MLP heads map the latent vector to the crystal orientation and phase class. Orientation and phase prediction tasks are optimized using a symmetry-aware geodesic loss \(L_{\mathrm{geo}}\) and a binary cross-entropy loss \(L_{\mathrm{cross}}\), respectively.  \textbf{b} Evolution of \(L_{\mathrm{geo}}\) and \(L_{\mathrm{cross}}\) during training. \textbf{c} A synthetic 4D-STEM dataset of a mixture of fcc $Cu$ and $Cu_{2}O$ crystal grains: (left) virtual bright-field image, where dark domains correspond to individual grains, and (right) a diffraction pattern randomly sampled from the scan position marked on the left. \textbf{d} Phase and orientation labels for the grains in (c). Phase labels include $Cu$, $Cu_{2}O$, and a background (BG) class, where BG corresponds to diffraction patterns recorded from scan regions without crystallites. For the grain highlighted on the left, a representative diffraction pattern and the corresponding Bragg disks (red) are shown on the right. \textbf{e} Model predictions of phase and orientation for the 4D-STEM data in (c). \textbf{f} Simulated patterns from the label (left) and from a representative prediction (right) for the grain highlighted in (d,e). Detected Bragg disks in (d) (red) are overlaid on simulated Bragg disks (black).}
\label{fig:FIG_6}
\end{center}
\end{figure}

To identify conditions under which the model predictions are spatially coherent and comparable to those from template matching, we examined the agreement between zone-axis directions predicted by the model across the scan grid and between the two methods. Here, we used the zone-axis directions rather than the full orientations because they lie in a common geometric space for both methods and are simpler to compare than full three-dimensional orientations. From this spatial coherence analysis, we identified scan regions in which the zone-axis directions predicted by the model are closely aligned across neighboring scan positions. We refer to these regions as highly correlated domains (Fig. 5d). Within these domains, we evaluated two-point spatial correlations by averaging dot products of zone-axis directions over pairs of scan positions. The model predictions show high two-point autocorrelation, as expected from the way these domains are identified. Notably, we observe comparably high autocorrelation for the template-matching predictions and high cross-correlation between the predictions from the two methods (Fig. 5e,f). Together, these results indicate that, in the spatially coherent domains, the model and template matching yield broadly similar orientation estimates. We observed consistent orientation mapping and correlation behavior across five independently trained models (Fig. S8,9). Based on these results, we suggest a potential route for utilizing model predictions for noisy experimental data, which is described in \textit{Supplementary Text S4}. We note that this observation is based on a single experimental dataset, and the extent to which it generalizes to other systems remains to be established.


\subsection*{Joint Prediction of Crystal Orientations and Phases from 4D-STEM Data}

Joint identification of crystal phase and orientation from 4D-STEM data provides direct access to microstructural descriptors that govern materials behavior, including phase distributions, interfacial character, and orientation relationships. These descriptors are central to understanding how material properties emerge and evolve during processing and application. For example, joint prediction can spatially resolve coexisting fcc $Cu$ and cuprous oxide $Cu_{2}O$ (space group $Pn\bar{3}m$, Fig. S1b) nanocrystals and their evolution within $Cu$-based electrocatalysts, which can impact $CO_{2} RR$ reaction pathways, catalytic selectivity, and activity\cite{cheng2025direct}. Here we show that our model can be extended to predict both crystal orientation and crystal phase from synthetic 4D-STEM data of a mixture of fcc $Cu$ and $Cu_{2}O$ nanocrystals.

To perform joint prediction, we added an additional MLP head to the model. This head maps the latent vector to a binary class label indicating whether a diffraction pattern corresponds to fcc $Cu$ or $Cu_{2}O$ crystal phase (Fig. 6a). Using a new set of simulated diffraction patterns from both crystals, the model was trained to predict both the binary phase label and the $3\times3$ rotation matrix representing the crystal orientation. During training, we optimized the phase prediction with the binary cross-entropy loss and the orientation prediction with the symmetry-aware geodesic loss. Both losses for training and validation data decreased and reached small values as the model learned to perform both tasks (Fig. 6b).

After training, we used the model to spatially resolve the crystal phase and orientation from synthetic 4D-STEM data of a mixture of fcc $Cu$ and $Cu_{2}O$ nanocrystals (Fig. 6c). In the mixture, the $Cu$ and $Cu_{2}O$ nanocrystals were present as individual grains with different orientations and bounded by the domain without crystallites (Fig. 6d). With our model, the accuracy of crystal phase prediction was $0.9973 \pm 0.0005$ (\textit{Methods}). The angular misalignment between symmetry-reduced crystallographic axes derived from the predictions and the labels was small for each axis (Fig. S10), with mean values of $0.037 \pm 0.079$ ($x$), $0.042 \pm 0.062$ ($y$), and $0.040 \pm 0.058$ ($z$) in radians. Overall, for a large fraction of grains, the phase and orientation maps from our model were broadly consistent with those from the labels (Fig. 6e). This result indicates that our model can jointly predict crystal phase and orientation with moderate accuracy for this synthetic 4D-STEM data. We observed comparable prediction accuracy across two independently trained models (Fig. S11).

However, the predictions were not accurate for some nanocrystals. Within some grains, the predicted phase and orientation varied across scan positions. These outcomes arise in part from the ambiguity that occurs when different phases or orientations yield nearly identical subsets of high-intensity Bragg disks (Fig. 6e, right). Both $Cu$ and $Cu_{2}O$ crystals have common cubic structures (Fig. S1), and the $Cu$ atoms in both materials are arranged in a fcc symmetry. Their lattice constants differ by less than 0.7 \AA{}. As such, diffraction patterns from the two phases can be highly similar for certain orientations. In cases where only a limited number of Bragg disks are present, such as in Fig. 6c, this similarity complicates reliable assignment of orientation, even for those experienced with diffraction indexing. Taken together, these results suggest that the model provides reasonable estimates of crystal phase and orientation from the 4D-STEM data, and that further improvements are needed to address the inaccuracies arising from this ambiguity.


\section*{Discussion}
Our transformer-based model enables rapid inference of crystalline microstructure from 4D-STEM diffraction data through direct prediction of structural attributes, including crystal orientation and phase. 
Unlike a few machine learning models that use 2D diffraction images to infer structural attributes \cite{zhu2024structural, munshi2022disentangling}, our model instead uses the set of Bragg disks.
As a result, we believe our model will be less sensitive to variations in imaging conditions, such as camera length and probe design, than the image-based approaches. When Bragg disks are reliably detected and their associated spatial frequencies fall within the range represented in the training data, a model trained for a given crystalline material can in principle be applied across multiple 4D-STEM datasets collected under different imaging conditions.
Additionally, we expect that processing a few Bragg disks will be more compute efficient than processing all of the pixels in an image.
Indeed, recently published concurrent work has performed inference on Bragg disks using an architecture combining convolutional layers and graph neural networks and has observed this to be a computationally efficient strategy~\cite{nathani2026accelerating}.
Beyond the structural attributes studied in this work, we anticipate that the model can be extended to predict additional crystallographic information, such as strain. Such extensions would complement orientation and phase information and provide a more detailed characterization of crystalline microstructure from 4D-STEM datasets.

Although the model achieves high accuracy on simulated diffraction patterns, its predictions yield lower cross-correlation for some noisy experimental 4D-STEM datasets. This discrepancy reflects a broader challenge in applying models trained on simulated data to experimental measurements. The statistical distributions of Bragg-disk features in experiments are not always identical to those in simulations because of substrate scattering, specimen thickness variation, and complex shape factors. These differences sometimes place limits on the performance of models solely trained on simulated data. Domain-confusion \cite{tzeng2014deep} and adversarial \cite{ganin2016domain} representation learning may be able to reduce the simulation–experiment domain gap in future work . When incorporated as auxiliary components within a supervised framework, these approaches may offer a promising route for improving the generalization of microstructure inference across diverse experimental conditions.

Beyond this limitation, a more fundamental challenge arises from the intrinsic similarity among diffraction patterns associated with different orientations and, in some cases, different phases. Multiple orientations can yield diffraction patterns that share the same subset of Bragg disks. When only a few disks are identifiable, assigning a single deterministic orientation to each diffraction pattern may not be well posed. This limitation is not specific to a particular inference method. Template matching can also assign similarly high correlation scores to templates from multiple orientations under these conditions. A probabilistic framework that provides a posterior distribution over orientations for a given pattern may offer a more faithful representation of the uncertainty inherent in noisy diffraction data. Future work should integrate model components that narrow the gap between simulated and experimental data and incorporate probabilistic reasoning to interpret structural information from diffraction patterns for robust inference of crystalline microstructure.


\section*{Methods}
\subsection*{Bragg Disk Embedding}

In our framework, a diffraction pattern is represented as a set of Bragg disk tokens, each described by three continuous features: radial distance, polar angle, and intensity. Accordingly, each diffraction pattern is represented as a $N\times 3$ matrix, where $N$ is the number of Bragg disks and $3$ is the feature dimension. Each feature is independently encoded, and the resulting encodings are summed to form a Bragg disk embedding.

To encode these features, we discretize the range of each feature into bins and map each bin to a vector with model dimension $d_{model}=384$. This procedure converts continuous features into vector representations for each Bragg disk within a diffraction pattern. 

Radial distance and intensity are encoded using an absolute positional encoding following Vaswani \cite{vaswani2017attention}. For a discretized feature bin index $b$, the $e$-th component of the corresponding $d_{\text{model}}$-dimensional embedding is

\begin{equation}
f(b,e;B) =
\begin{cases}
\sin\left( \dfrac{b}{B^{\frac{2e}{d_{\text{model}}}}} \right), & \text{if } e \text{ is even} \\
\cos\left( \dfrac{b}{B^{\frac{2e}{d_{\text{model}}}}} \right), & \text{if } e \text{ is odd}
\end{cases}
\quad \text{for } e = 0, 1, \ldots, d_{\text{model}} - 1,
\end{equation}
where $B$ is a feature-specific denominator. 
The radial distance in the range $[0,2.99] \mathring{\mathrm{A}}^{-1}$ 
is discretized into 256 bins, and the corresponding bin index $k_r$ is encoded using $f(k_r,e;B_r)$ with $B_r = 5000$. 
The intensity in the range $[0.001,1]$ is discretized into 64 bins, and the corresponding bin index $I$ is encoded using $f(I,e;B_I)$ with $B_I = 10000$.

The polar angle in the range $[-\pi,\pi)$ is discretized into 360 bins. For each polar angle bin $k_{\theta}$, we use the sinusoidal encoding reported by Ning \cite{ning2020polar}:

\begin{equation}
    f_{\textit{k}_{\theta}}(\textit{k}_{\theta}) = \mathbf{c}_{k_{\theta},0} + \frac{1}{15}\sum_{i=1}^{15} \left( \mathbf{a}_{k_{\theta},i} \cos(ik_{\theta}) + \mathbf{b}_{k_{\theta},i} \sin(ik_{\theta}) \right)
\end{equation}
where $\mathbf{c}_{k_{\theta},0}$, $\mathbf{a}_{k_{\theta},i}$, and $\mathbf{b}_{k_{\theta},i}$ are learnable vectors with dimension $d_{model}$. The prefactor $\frac{1}{15}$ normalizes the contribution of the sinusoidal terms. The polar angle coordinate is periodic and requires an encoding that preserves continuity across the $-\pi$ to $\pi$ boundary. This sinusoidal encoding enforces this requirement such that angular directions on either side of the boundary are mapped to similar encodings. Standard absolute positional encoding, which treats the angular range as linear, assigns distant representations to physically adjacent directions and is therefore not used for polar angle encoding. 

After encoding, each diffraction pattern is represented as a tensor of size  $N\times 3 \times d_{model}$. Summing the feature encodings for each Bragg disk yields a Bragg disk embedding matrix of size $N\times d_{model}$, which serves as the model input. Because diffraction patterns contain varying numbers of Bragg disks, [PAD] tokens and corresponding masks are used to ensure a consistent input length.

\subsection*{The symmetry-aware geodesic loss $L_{\mathrm{geo}}$}

To train the model for orientation prediction, we employ a symmetry-aware geodesic loss that accounts for the proper point-group symmetries of the crystal, denoted by  $G$. This loss formulation ensures that orientations related by symmetry operations $g \in G$, which produce indistinguishable diffraction patterns, are treated as physically equivalent during training.

After model prediction, we generate the full set of symmetry-equivalent orientations by applying every proper point-group operation $g$ to the orientation label $\{\, g R_{\mathrm{label}} \;|\; g \in G \,\}$ (\textit{Table S1}). Here, the symmetry operators act on the left of $R_{\mathrm{label}}$ because the orientations used in this work follow the convention adopted in the py4DSTEM simulations and analysis. Under this convention, reciprocal lattice vectors are transformed from the crystal frame to the laboratory frame by the transpose of the corresponding rotation matrix \cite{savitzky2021py4dstem}. We compare the predicted rotation matrix $R_{\mathrm{pred}}$ with each symmetry-equivalent orientation $g R_{\mathrm{label}}$ using the geodesic distance on $\mathrm{SO}(3)$:
\[
d\!\left(R_{\mathrm{pred}}, g R_{\mathrm{label}}\right)
= \arccos\!\left(
\frac{\mathrm{tr}\!\left(R_{\mathrm{pred}} (g R_{\mathrm{label}})^{\top}\right) - 1}{2}
\right),
\]
where $\mathrm{tr}(\cdot)$ denotes the matrix trace. The geodesic distance represents the smallest rotation angle required to align the two rotation matrices and is expressed in radians. The symmetry-aware geodesic loss for a single diffraction pattern $L_{\mathrm{geo}}$ is then defined as the minimum of these distances:
\[
L_{\mathrm{geo}}\!\left(R_{\mathrm{pred}}, R_{\mathrm{label}}\right)
= \min_{g \in G} d\!\left(R_{\mathrm{pred}}, g R_{\mathrm{label}}\right).
\]

With the symmetry-aware loss, the model can map a diffraction pattern to any one of its symmetry-equivalent orientations. This behavior differs from template matching, where each symmetry-equivalent set is typically reduced to a single representative orientation and every pattern is assigned only to that representative. Although the symmetry reduction is convenient for indexing, it can introduce instability when used in a predictive model. Under the symmetry-reduction framework, a prediction may lie close to a symmetry-equivalent orientation while remaining far from the chosen representative, which can lead to an artificially large geodesic distance. This overestimation of orientation error can result in large fluctuations in the training loss. In our framework, the model is penalized only for deviations from the nearest symmetry-equivalent orientation, which prevents inflated errors and stabilizes optimization.

\subsection*{Model Architecture}
The model consists of two modules: a transformer encoder \cite{vaswani2017attention} and task-specific MLP heads (Fig. 1). The transformer encoder takes the Bragg disk embedding matrix of size $N \times d_{model}$ as input and processes it with three encoder blocks, each comprising multi-head self-attention with eight heads and a feed-forward network. These layers capture contextual relationships among Bragg disks and update their representations accordingly. The output of the transformer encoder is a contextual embedding matrix of size $N \times d_{model}$. Mean pooling over the token dimension produces a latent vector of size $d_{model}$ representing the diffraction pattern.

The MLP heads map this latent vector to crystallographic information. For orientation prediction, an MLP outputs a 9-dimensional vector that is reshaped into a $3 \times 3$ matrix and further normalized into a proper rotation matrix using the approach of Levinson \textit{et al.} \cite{levinson2020analysis}. For crystal phase prediction, a separate MLP maps the latent vector to a single binary label (Fig. 6). All models were implemented in PyTorch \cite{paszke2019pytorch}. Training and inference were performed using PyTorch versions 2.8.0 and 2.9.0, depending on software availability across computing systems; no qualitative differences in model behaviour were observed. More details of training are provided in \textit{Supplementary Text S5}.

\subsection*{Synthetic data generation for training, validation, and test datasets}

All synthetic datasets were generated using dynamical diffraction simulations based on the Bloch-wave formalism \cite{de2003introduction,zeltmann2023uncovering}, as implemented in py4DSTEM (version 0.14.08) \cite{savitzky2021py4dstem}. In these simulations, the crystal orientation and thickness define the diffraction conditions. For the training and validation datasets, we sampled symmetrically unique zone-axis directions on the unit sphere using a spherical linear interpolation formula \cite{shoemake1985animating,ophus2022automated}. For each zone-axis direction, we then sampled in-plane rotation angles and crystal thicknesses. The crystal orientation was obtained by aligning the zone-axis with the incident electron beam direction and subsequently applying the in-plane rotation about that axis. Details of the sampling are provided in \textit{Supplementary Text S2}. After simulation, we identified cases in which different combinations of crystal orientation and thickness produced indistinguishable diffraction patterns, and retained a single representative orientation label for each such pattern. This procedure avoids an ill-posed learning problem in which the model would otherwise be required to map the same diffraction pattern to multiple target orientations. The simulated dataset was randomly divided into training and validation sets using an 80/20 split. After training, the model with the lowest validation loss was selected for evaluation on the test dataset and 4D-STEM datasets.

To generate the test dataset, we independently and randomly sampled zone-axis directions, in-plane rotation angles, and crystal thicknesses, and generated the corresponding diffraction patterns using dynamical diffraction simulations. For the model trained to predict orientations from fcc $Cu$ diffraction patterns, the dataset contained 16,797,888 diffraction patterns for training, 4,199,472 diffraction patterns for validation, and 65,536 diffraction patterns for testing. For the model trained to jointly predict crystal orientations and phases, the dataset contained 33,621,696 diffraction patterns for training and 8,405,424 diffraction patterns for validation. The test data for the joint prediction correspond to the synthetic 4D-STEM dataset shown in Fig. 6. We note that the dynamical diffraction simulations output a list of Bragg disks with 2D positions and intensities for each diffraction pattern, rather than 2D diffraction images. For model training and prediction, the intensities of Bragg disks within each diffraction pattern were normalized by the maximum disk intensity in that pattern.

\subsection*{Mapping a 4D-STEM diffraction pattern to a set of Bragg disks}

The mapping of a diffraction pattern image to a set of Bragg disks involves image pre-processing and template matching. In the pre-processing step, each 4D-STEM diffraction pattern image was centered and the diffuse background was removed using a difference-of-Gaussians procedure \cite{lowe2004distinctive}. From the processed image, the Bragg disks are identified using a template matching approach \cite{savitzky2021py4dstem}.
This approach differs from template matching used for orientation identification. It measures the cross-correlation of the vacuum probe template with a pre-processed image and locates Bragg disk positions by identifying local maxima of the cross-correlation. In our work, the vacuum probe was taken as the direct-beam profile averaged over scan regions without crystallites.

\subsection*{Generation of a synthetic 4D-STEM dataset of fcc $Cu$ and $Cu_{2}O$ crystals}

The synthetic 4D-STEM dataset for the polycrystalline mixture was generated through a sequence of steps. A 2D scan grid of $128 \times 128$ positions was first divided into 60 spatial domains, each representing an individual crystal grain. Equal numbers of domains were then randomly assigned $Cu$ fcc and $Cu_{2}O$ phase labels, and a crystal orientation was randomly sampled for each domain. Based on the assigned crystal phase and orientation, dynamical diffraction patterns were simulated. To incorporate experimentally realistic diffraction signals into simulated data, background diffraction patterns recorded from regions without crystallites were sampled, and a direct-beam template was extracted from the diffraction data. The simulated diffraction patterns were convolved with the direct-beam template to produce 2D images with sparse disk distributions. The resulting images were then linearly combined with the background diffraction patterns to generate the final synthetic 4D-STEM diffraction patterns.

For the dynamical diffraction simulations, a fixed crystal thickness of 21 \AA{} was used. The background diffraction patterns were obtained from the 4D-STEM measurements described in the subsection titled \textit{Experimental 4D-STEM measurements of $Cu$ dendritic crystals} in the Methods section.

\subsection*{Calculation of sparse correlation scores and scan-grid correlation functions}

Sparse correlation scores were calculated following the implementation provided in the py4DSTEM software package \cite{ophus2022automated}. Unless otherwise specified, default parameters were used for the sparse correlation calculations. Cross-correlation scores $Q$ were computed following the method described in Refs. \cite{cautaerts2022free,folastre2024improved}:
\begin{equation}
    Q = \frac{\sum_{p=1} T(x_{p},y_{p}) D(x_{p},y_{p})}{\sqrt{\sum_{p=1}T(x_{p},y_{p})}\sqrt{\sum_{p=1}D(x_{p},y_{p})}}
\end{equation}
where $T(x_{p},y_{p})$ and $D(x_{p},y_{p})$ denote the intensity values of pixel $p$ in a simulated template pattern and an experimental diffraction pattern, respectively. The simulated template patterns were generated using kinematic diffraction simulations. Before computing $Q$, the intensity values of the experimental diffraction patterns were transformed using a gamma correction with an exponent of 0.5 \cite{jeong2021automated,cautaerts2022free}. The direct beam in the template pattern was excluded from the calculation of $Q$.

To identify spatially correlated predictions across the scan grid, we first reduced the predicted crystal orientations to their corresponding zone-axis directions. For each scan position, we computed the dot product between its zone-axis direction and those of its 24 nearest-neighboring scan positions, excluding neighbors without assigned zone-axis directions. The average of these dot products was calculated, and if the average dot product was larger than 0.98, the scan position was assigned to a highly correlated domain. The two-point auto-correlation function was computed by averaging the dot products between zone-axis directions assigned to pairs of scan positions within the scan grid. The two-point cross-correlation function was computed by averaging the dot products between a zone-axis direction obtained from the model prediction at one scan position and a zone-axis direction obtained from template matching at another scan position.

\subsection*{Experimental 4D-STEM measurements of $Cu$ dendritic crystals}
A detailed description of the 4D-STEM measurements is provided in Ref.~\cite{kim2025operando}. The measurements used in this work were performed using 4D-STEM with an electrochemical liquid-cell scanning transmission electron microscope operated at an accelerating voltage of 300 keV. Diffraction data were acquired using an electron microscope pixel array detector under low-dose conditions. The effective liquid thickness was reduced by electrochemically generating hydrogen bubbles, yielding a thin liquid layer with an effective thickness of approximately 120~nm \cite{kim2025operando}. Measurements were conducted at temperatures down to  $-40,^{\circ}\mathrm{C}$ for durations of up to $\sim$1~h using a total electron dose of $\sim$40~e$^{-}$/\AA, where no noticeable beam-induced damage was observed after acquisition.

\subsection*{Benchmarking computation time}

To benchmark computation time for orientation mapping, the experiments for model prediction and template matching were executed on a desktop computer equipped with an Intel® Core™ i9-14900K processor with 32 CPU cores, an NVIDIA GeForce RTX™ 4070 Ti SUPER GPU, and 64 GB of RAM, running the Arch Linux operating system. For these benchmarks, the preprocessed 4D-STEM inputs were synthetically constructed by assigning to each scan position a set of Bragg disks corresponding to a simulated diffraction pattern, randomly sampled from the training and validation datasets. Template matching was implemented using the py4DSTEM software package \cite{ophus2019four,savitzky2021py4dstem} and executed on CPUs. Model predictions were performed in both CPU-only mode and GPU-accelerated mode.

All benchmarks were conducted through a command-line interface to minimize interference from graphical user interface activity and background processes. Each benchmark was repeated ten times sequentially to collect statistics. As a result of the sequential runs, small variations arising from system-level caching effects cannot be fully excluded. The codes were allowed to freely utilize available CPU cores through multithreading. The template matching relies on a NumPy backend but does not heavily involve matrix multiplications between large matrices or tensors. Consequently, it makes limited use of multithreading and utilization of multiple CPU cores. In CPU-only mode, model prediction used multiple CPU cores, with no explicit constraints imposed on multithreading. In GPU-accelerated mode, model prediction used on average a single CPU core, with the primary computation performed on the GPU. 

For template matching, orientation mapping was performed using the default parameters provided by py4DSTEM. Depending on the choice of parameters, the size of the template library and the computation time can vary. Model prediction involves four subprocesses: loading the pre-processed 4D-STEM diffraction data, transformation from Cartesian to polar coordinates and construction of the model input tensors, orientation prediction using trained models, and mapping of the predicted orientations back to the corresponding scan positions. Computation time for both template matching and model prediction was measured using wall-clock elapsed time reported by the Linux operating system. For model prediction, wall-clock time for each subprocess was additionally recorded separately. A complete summary of timing results are provided in \textit{Supplementary Table S2-4}.

\subsection*{Evaluating the joint prediction of crystallographic orientation and phase}

Prediction accuracy for the binary phase classification was evaluated by comparing the predicted phase labels with the corresponding ground-truth labels at each scan position. Let $y_i \in \{0,1\}$ denote the true class label and $\hat{y}_i \in \{0,1\}$ the predicted label for the $i$-th sample, with a total of $N$ samples. The prediction accuracy $p$ was computed as

\begin{equation}
p = \frac{1}{N} \sum_{i=1}^{N} \mathbb{I}(y_i = \hat{y}_i),
\end{equation}
where $\mathbb{I}(\cdot)$ is the indicator function, equal to 1 when the prediction is correct and 0 otherwise. To quantify the statistical uncertainty of the accuracy estimate, each prediction outcome was modeled as an independent Bernoulli trial (correct or incorrect). Under this assumption, the standard deviation of the accuracy estimate is given by

\begin{equation}
\sigma_p = \sqrt{\frac{p(1 - p)}{N}}.
\end{equation}
The reported accuracy values are expressed as $p \pm \sigma_p$.

To assess the accuracy of the orientation predictions, we computed the angular difference between the symmetry-reduced crystallographic axes derived from the label and predicted orientations. Symmetry reduction was performed using the full crystallographic point group, as implemented in py4DSTEM. We did not use geodesic loss for a direct comparison between predicted and label orientations. This is because the synthetic 4D-STEM data is generated from randomly sampled orientations, which can lead to high geodesic loss even for reasonably accurate predictions (Fig. 2d-g); the label orientations in the data are sometimes not symmetrically equivalent to those used in training, occasionally resulting in inflated geodesic loss.

\section*{Data availability}
Data are provided within the manuscript or supplementary information files. The code for generating synthetic training data is provided in the corresponding section. Any additional raw data that support the findings in this paper are available upon reasonable request.

\section*{Code availability}
The transformer-based orientation mapping framework developed in this work is available as an open-source Python package at https://github.com/thiedelab/microstructureInference. The package includes codes for generating synthetic training, validation, and test data.

\section*{Acknowledgements}
K.J. thanks Chia-Hao Lee and Chuqiao Shi for helpful discussions. K.J. and E.K are supported by the Eric and Wendy Schmidt AI in Science Postdoctoral Fellowship, a program of Schmidt Sciences, LLC. This work used Delta and DeltaAI at the National Center for Supercomputing Applications through allocation CIS240889 from the Advanced Cyberinfrastructure Coordination Ecosystem: Services $\&$ Support (ACCESS) program, which is supported by U.S. National Science Foundation grants $\#$2138259, $\#$2138286, $\#$2138307, $\#$2137603, and $\#$2138296 \cite{boerner2023access}. This research was also supported in part through computational resources and services supported by the Cornell University Center for Advanced Computing, which receives funding from Cornell University, the National Science Foundation, and members of its Partner Program.

\clearpage

\section*{Author information}
\subsection*{Authors and Affiliations}

\textbf{Department of Chemistry and Chemical Biology, Baker Lab, Cornell University, Ithaca, New York 14853, United States}
\newline
Kwanghwi Je, Ellis R. Kennedy, Sung-in Kim, Yao Yang, Erik H. Thiede

\subsection*{Contributions}
K.J. and E.T. conceived the project. K.J. developed the methodology and performed the analysis, with support from E.K.; S.K. and Y.Y provided experimental four dimensional scanning transmission electron microscopy data. All authors contributed to the discussion and revision of the manuscript. E.T. supervised the project.

\subsection*{Corresponding author}

Correspondence to Erik H. Thiede.

\section*{Ethics declarations}

\subsection*{Competing interests}
The authors declare no competing interests.


\bibliography{main}

@article{morawiec2007algorithm,
  title={An algorithm for refinement of lattice parameters using CBED patterns},
  author={Morawiec, A},
  journal={Ultramicroscopy},
  volume={107},
  number={4-5},
  pages={390--395},
  year={2007},
  publisher={Elsevier}
}

@article{rauch2010automated,
  title={Automated nanocrystal orientation and phase mapping in the transmission electron microscope on the basis of precession electron diffraction},
  author={Rauch, Edgar F and Portillo, Joaquin and Nicolopoulos, Stavros and Bultreys, Daniel and Rouvimov, Sergei and Moeck, Peter},
  journal={Z. Krist.},
  year={2010}
}

@article{geng2024grain,
  title={Grain boundary engineering for efficient and durable electrocatalysis},
  author={Geng, Xin and Vega-Paredes, Miquel and Wang, Zhenyu and Ophus, Colin and Lu, Pengfei and Ma, Yan and Zhang, Siyuan and Scheu, Christina and Liebscher, Christian H and Gault, Baptiste},
  journal={Nat. Commun.},
  volume={15},
  number={1},
  pages={8534},
  year={2024},
  publisher={Nature Publishing Group UK London}
}

@article{xu2020charge,
  title={Charge distribution guided by grain crystallographic orientations in polycrystalline battery materials},
  author={Xu, Zhengrui and Jiang, Zhisen and Kuai, Chunguang and Xu, Rong and Qin, Changdong and Zhang, Yan and Rahman, Muhammad Mominur and Wei, Chenxi and Nordlund, Dennis and Sun, Cheng-Jun and others},
  journal={Nat. Commun.},
  volume={11},
  number={1},
  pages={83},
  year={2020},
  publisher={Nature Publishing Group UK London}
}

@article{bishara2021understanding,
  title={Understanding grain boundary electrical resistivity in {Cu}: the effect of boundary structure},
  author={Bishara, Hanna and Lee, Subin and Brink, Tobias and Ghidelli, Matteo and Dehm, Gerhard},
  journal={ACS nano},
  volume={15},
  number={10},
  pages={16607--16615},
  year={2021},
  publisher={ACS Publications}
}

@article{weaver1990diffusivity,
  title={Diffusivity of ultrasound in polycrystals},
  author={Weaver, Richard L},
  journal={J. Mech. Phys. Solids},
  volume={38},
  number={1},
  pages={55--86},
  year={1990},
  publisher={Elsevier}
}

@article{kim2017phonon,
  title={Phonon scattering by dislocations at grain boundaries in polycrystalline {Bi\textsubscript{0.5}Sb\textsubscript{1.5}Te\textsubscript{3}}},
  author={Kim, Hyun-Sik and Kim, Sang Il and Lee, Kyu Hyoung and Kim, Sung Wng and Snyder, G Jeffrey},
  journal={Phys. Status Solidi B},
  volume={254},
  number={5},
  pages={1600103},
  year={2017},
  publisher={Wiley Online Library}
}

@article{bauer2014high,
  title={High-strength cellular ceramic composites with {3D} microarchitecture},
  author={Bauer, Jens and Hengsbach, Stefan and Tesari, Iwiza and Schwaiger, Ruth and Kraft, Oliver},
  journal={Proc. Natl. Acad. Sci. USA},
  volume={111},
  number={7},
  pages={2453--2458},
  year={2014},
  publisher={National Academy of Sciences}
}

@article{fang2011revealing,
  title={Revealing extraordinary intrinsic tensile plasticity in gradient nano-grained copper},
  author={Fang, TH and Li, WL and Tao, NR and Lu, K},
  journal={Science},
  volume={331},
  number={6024},
  pages={1587--1590},
  year={2011},
  publisher={American Association for the Advancement of Science}
}

@article{zhou2018enhanced,
  title={Enhanced thermal stability of nanograined metals below a critical grain size},
  author={Zhou, X and Li, XY and Lu, K},
  journal={Science},
  volume={360},
  number={6388},
  pages={526--530},
  year={2018},
  publisher={American Association for the Advancement of Science}
}

@article{cheng2018extra,
  title={Extra strengthening and work hardening in gradient nanotwinned metals},
  author={Cheng, Zhao and Zhou, Haofei and Lu, Qiuhong and Gao, Huajian and Lu, Lei},
  journal={Science},
  volume={362},
  number={6414},
  pages={eaau1925},
  year={2018},
  publisher={American Association for the Advancement of Science}
}

@article{lowe2004distinctive,
  title={Distinctive image features from scale-invariant keypoints},
  author={Lowe, David G},
  journal={Int. J. Comput. Vis.},
  volume={60},
  number={2},
  pages={91--110},
  year={2004},
  publisher={Springer}
}

@book{de2003introduction,
  title={Introduction to conventional transmission electron microscopy},
  author={De Graef, Marc},
  year={2003},
  publisher={Cambridge university press}
}

@article{zeltmann2023uncovering,
  title={Uncovering polar vortex structures by inversion of multiple scattering with a stacked Bloch wave model},
  author={Zeltmann, Steven E and Hsu, Shang-Lin and Brown, Hamish G and Susarla, Sandhya and Ramesh, Ramamoorthy and Minor, Andrew M and Ophus, Colin},
  journal={Ultramicroscopy},
  volume={250},
  pages={113732},
  year={2023},
  publisher={Elsevier}
}

@article{paszke2019pytorch,
  title={Pytorch: An imperative style, high-performance deep learning library},
  author={Paszke, Adam and Gross, Sam and Massa, Francisco and Lerer, Adam and Bradbury, James and Chanan, Gregory and Killeen, Trevor and Lin, Zeming and Gimelshein, Natalia and Antiga, Luca and others},
  journal={Advances in neural information processing systems},
  volume={32},
  year={2019}
}

@article{zhu2024structural,
  title={Structural degeneracy and formation of crystallographic domains in epitaxial LaFeO3 films revealed by machine-learning assisted {4D-STEM}},
  author={Zhu, Menglin and Lanier, Joseph and Flores, Jose and da Cruz Pinha Barbosa, Victor and Russell, Daniel and Haight, Becky and Woodward, Patrick M and Yang, Fengyuan and Hwang, Jinwoo},
  journal={Sci. Rep.},
  volume={14},
  number={1},
  pages={4198},
  year={2024},
  publisher={Nature Publishing Group UK London}
}

@article{munshi2022disentangling,
  title={Disentangling multiple scattering with deep learning: application to strain mapping from electron diffraction patterns},
  author={Munshi, Joydeep and Rakowski, Alexander and Savitzky, Benjamin H and Zeltmann, Steven E and Ciston, Jim and Henderson, Matthew and Cholia, Shreyas and Minor, Andrew M and Chan, Maria KY and Ophus, Colin},
  journal={npj Comput. Mater.},
  volume={8},
  number={1},
  pages={254},
  year={2022},
  publisher={Nature Publishing Group UK London}
}

@article{folastre2024improved,
  title={Improved {ACOM} pattern matching in {4D-STEM} through adaptive sub-pixel peak detection and image reconstruction},
  author={Folastre, Nicolas and Cao, Junhao and Oney, Gozde and Park, Sunkyu and Jamali, Arash and Masquelier, Christian and Croguennec, Laurence and Veron, Muriel and Rauch, Edgar F and Demorti{\`e}re, Arnaud},
  journal={Sci. Rep.},
  volume={14},
  number={1},
  pages={12385},
  year={2024},
  publisher={Nature Publishing Group UK London}
}

@article{rauch2014automated,
  title={Automated crystal orientation and phase mapping in {TEM}},
  author={Rauch, EF and V{\'e}ron, MJMC},
  journal={Mater. Charact.},
  volume={98},
  pages={1--9},
  year={2014},
  publisher={Elsevier}
}

@inproceedings{ning2020polar,
  title={Polar Relative Positional Encoding for Video-Language Segmentation},
  author={Ning, Ke and Xie, Lingxi and Wu, Fei and Tian, Qi},
  booktitle={IJCAI},
  volume={9},
  pages={10},
  year={2020}
}

@article{levinson2020analysis,
  title={An analysis of svd for deep rotation estimation},
  author={Levinson, Jake and Esteves, Carlos and Chen, Kefan and Snavely, Noah and Kanazawa, Angjoo and Rostamizadeh, Afshin and Makadia, Ameesh},
  journal={Adv. Neural Inf. Process. Syst.},
  volume={33},
  pages={22554--22565},
  year={2020}
}

@article{ophus2022automated,
  title={Automated crystal orientation mapping in {py4DSTEM} using sparse correlation matching},
  author={Ophus, Colin and Zeltmann, Steven E and Bruefach, Alexandra and Rakowski, Alexander and Savitzky, Benjamin H and Minor, Andrew M and Scott, Mary C},
  journal={Microsc. Microanal.},
  volume={28},
  number={2},
  pages={390--403},
  year={2022},
  publisher={Cambridge University Press}
}

@article{savitzky2021py4dstem,
  title={{py4DSTEM}: A software package for four-dimensional scanning transmission electron microscopy data analysis},
  author={Savitzky, Benjamin H and Zeltmann, Steven E and Hughes, Lauren A and Brown, Hamish G and Zhao, Shiteng and Pelz, Philipp M and Pekin, Thomas C and Barnard, Edward S and Donohue, Jennifer and DaCosta, Luis Rangel and others},
  journal={Microsc. Microanal.},
  volume={27},
  number={4},
  pages={712--743},
  year={2021},
  publisher={Cambridge University Press}
}

@article{ophus2019four,
  title={Four-dimensional scanning transmission electron microscopy ({4D-STEM}): From scanning nanodiffraction to ptychography and beyond},
  author={Ophus, Colin},
  journal={Microsc. Microanal.},
  volume={25},
  number={3},
  pages={563--582},
  year={2019},
  publisher={Cambridge University Press}
}

@article{jeong2021automated,
  title={Automated crystal orientation mapping by precession electron diffraction-assisted four-dimensional scanning transmission electron microscopy using a scintillator-based CMOS detector},
  author={Jeong, Jiwon and Cautaerts, Niels and Dehm, Gerhard and Liebscher, Christian H},
  journal={Microsc. Microanal.},
  volume={27},
  number={5},
  pages={1102--1112},
  year={2021},
  publisher={Oxford University Press}
}

@article{cautaerts2022free,
  title={Free, flexible and fast: Orientation mapping using the multi-core and GPU-accelerated template matching capabilities in the Python-based open source {4D-STEM} analysis toolbox Pyxem},
  author={Cautaerts, Niels and Crout, Phillip and {\AA}nes, H{\aa}kon W and Prestat, Eric and Jeong, Jiwon and Dehm, Gerhard and Liebscher, Christian H},
  journal={Ultramicroscopy},
  volume={237},
  pages={113517},
  year={2022},
  publisher={Elsevier}
}

@article{liu20244d,
  title={{4D-STEM} mapping of nanocrystal reaction dynamics and heterogeneity in a graphene liquid cell},
  author={Liu, Chang and Lin, Oliver and Pidaparthy, Saran and Ni, Haoyang and Lyu, Zhiheng and Zuo, Jian-Min and Chen, Qian},
  journal={Nano Lett.},
  volume={24},
  number={13},
  pages={3890--3897},
  year={2024},
  publisher={ACS Publications}
}

@article{cheng2025direct,
  title={Direct Visualization and Quantitative Insights into the Formation and Phase Evolution of {Cu} Nanoparticles via In Situ Liquid Phase {4D-STEM}},
  author={Cheng, Ningyan and Sun, Hongyu and Pivak, Yevheniy and Liebscher, Christian H},
  journal={Adv. Sci.},
  volume={12},
  number={19},
  pages={2500706},
  year={2025},
  publisher={Wiley Online Library}
}

@article{vaswani2017attention,
  title={Attention is all you need},
  author={Vaswani, Ashish and Shazeer, Noam and Parmar, Niki and Uszkoreit, Jakob and Jones, Llion and Gomez, Aidan N and Kaiser, {\L}ukasz and Polosukhin, Illia},
  journal={Adv. Neural Inf. Process. Syst.},
  volume={30},
  year={2017}
}

@article{kim2025operando,
author = {Kim, Sungin and Briega-Martos, Valentin and Liu, Shikai and Je, Kwanghwi and Shi, Chuqiao and Stephens, Katherine Marusak and Zeltmann, Steven E. and Zhang, Zhijing and Guzman-Soriano, Rafael and Li, Wenqi and Jiang, Jiahong and Choi, Juhyung and Negash, Yafet J. and Walden, Franklin S. II and Marthe, Nelson L. Jr. and Wellborn, Patrick S. and Guo, Yaofeng and Damiano, John and Han, Yimo and Thiede, Erik H. and Yang, Yao},
title = {Operando Heating and Cooling Electrochemical {4D-STEM} Probing Nanoscale Dynamics at Solid–Liquid Interfaces},
journal = {J. Am. Chem. Soc.},
volume = {147},
number = {27},
pages = {23654-23671},
year = {2025},
}

@inproceedings{shoemake1985animating,
  title={Animating rotation with quaternion curves},
  author={Shoemake, Ken},
  booktitle={Proceedings of the 12th annual conference on Computer graphics and interactive techniques},
  pages={245--254},
  year={1985}
}

@inproceedings{boerner2023access,
author = {Boerner, Timothy J. and Deems, Stephen and Furlani, Thomas R. and Knuth, Shelley L. and Towns, John},
title = {ACCESS: Advancing Innovation: {NSF}’s Advanced Cyberinfrastructure Coordination Ecosystem: Services \& Support},
year = {2023},
isbn = {9781450399852},
publisher = {Association for Computing Machinery},
address = {New York, NY, USA},
booktitle = {Practice and Experience in Advanced Research Computing 2023: Computing for the Common Good},
pages = {173–176},
numpages = {4},
location = {Portland, OR, USA},
series = {PEARC '23}
}

@article{belsky2002new,
  title={New developments in the Inorganic Crystal Structure Database (ICSD): accessibility in support of materials research and design},
  author={Belsky, Alec and Hellenbrandt, Mariette and Karen, Vicky Lynn and Luksch, Peter},
  journal={Acta Crystallogr. Sec. B Struct. Sci.},
  volume={58},
  number={3},
  pages={364--369},
  year={2002},
  publisher={International Union of Crystallography}
}

@misc{tzeng2014deep,
  author = {Tzeng, Eric and Hoffman, Judy and Zhang, Ning and Saenko, Kate and Darrell, Trevor},
  title  = {Deep domain confusion: maximizing for domain invariance},
  howpublished = {Preprint at \url{https://doi.org/10.48550/arXiv.1412.3474}},
  year   = {2014}
}

@article{ganin2016domain,
  title={Domain-adversarial training of neural networks},
  author={Ganin, Yaroslav and Ustinova, Evgeniya and Ajakan, Hana and Germain, Pascal and Larochelle, Hugo and Laviolette, Fran{\c{c}}ois and March, Mario and Lempitsky, Victor},
  journal={J. Mach. Learn. Res.},
  volume={17},
  number={59},
  pages={1--35},
  year={2016}
}

@inproceedings{marcel2010torchvision,
  title={Torchvision the machine-vision package of torch},
  author={Marcel, S{\'e}bastien and Rodriguez, Yann},
  booktitle={Proceedings of the 18th ACM international conference on Multimedia},
  pages={1485--1488},
  year={2010}
}

@misc{loshchilov2017decoupled,
  author = {Loshchilov, Ilya and Hutter, Frank},
  title  = {Decoupled weight decay regularization},
  howpublished = {Preprint at \url{https://doi.org/10.48550/arXiv.1711.05101}},
  year   = {2017}
}

@article{nathani2026accelerating,
  title={Accelerating electron diffraction analysis using graph neural networks and attention mechanisms},
  author={Nathani, Anvesh and McCray, Arthur RC and Liu, Yingtao and Ding, Hanping and Kazempoor, Pejman and Xu, Shuozhi and Ophus, Colin and Ghamarian, Iman},
  journal={npj Comput. Mater.},
  year={2026},
  publisher={Nature Publishing Group UK London}

}

@article{aran2020role,
  title={The role of in situ generated morphological motifs and {Cu(I)} species in {C\textsubscript{2+}} product selectivity during {CO\textsubscript{2}} pulsed electroreduction},
  author={Ar{\'a}n-Ais, Rosa M and Scholten, Fabian and Kunze, Sebastian and Rizo, Rub{\'e}n and Roldan Cuenya, Beatriz},
  journal={Nat. Energy},
  volume={5},
  number={4},
  pages={317--325},
  year={2020},
  publisher={Nature Publishing Group UK London}
}
\bibliographystyle{naturemag}

\end{document}


\maketitle
\vspace{-1.5cm}
\newcommand{\TOCentry}[3]{%
\hspace{1.5em}\textbf{#1}\;
\parbox[t]{0.70\textwidth}{#2} & #3 \\}

\noindent\textbf{Table of Contents}

\begin{center}
\begin{tabular}{@{}p{0.85\textwidth}r@{}}
\textbf{Supplementary Texts S1-S5} & {2 - 6} \\
\TOCentry{S1}{Contextual Learning and Attention Patterns in the Transformer Encoder}{2}
\noalign{\vskip 0.5em}
\TOCentry{S2}{Orientation sampling and simulation for generating training and validation diffraction datasets}{4}
\noalign{\vskip 0.5em}
\TOCentry{S3}{Data augmentation of Bragg disk features}{5}
\noalign{\vskip 0.5em}
\TOCentry{S4}{A potential route for utilizing model predictions on noisy experimental 4D-STEM data}{5}
\noalign{\vskip 0.5em}
\TOCentry{S5}{Training configuration and loss convergence}{6}
\noalign{\vskip 0.5em}
\textbf{Supplementary Figures S1-S20} & {7 - 25}\\
\noalign{\vskip 0.5em}
\textbf{Supplementary Tables S1-S4} & {26 - 29}\\
\noalign{\vskip 0.5em}
\textbf{Legends for Movie S1} & {30} \\
\noalign{\vskip 0.5em}
\textbf{References} & {31} \\
\end{tabular}
\end{center}

\hspace{1.0cm}

\noindent\textbf{Other Supplementary Materials for this manuscript include the following: }
\begin{center}
\begin{tabular}{@{}p{0.78\textwidth}r@{}}
\TOCentry{}{Supplementary Movie 1}{}
\end{tabular}
\end{center}

\clearpage

\section*{Supplementary Text S1: Contextual Learning and Attention Patterns in the Transformer Encoder}

To gain preliminary insight into contextual learning in the transformer-based model, we examined the attention weights produced by the trained transformer encoder at different layers \cite{vaswani2017attention}. The model was trained to map Bragg disk features (positions and intensities) in a diffraction pattern to the crystallographic orientation of a single crystal. The analysis here focuses on the layer-wise attention behavior within the encoder of the model. While neural networks remain fundamentally difficult to interpret and any conclusions must be treated with caution, several empirical trends emerged that may help to contextualize how the model organizes information across layers.

A recurring pattern in the lowest encoder layer, seen in diffraction data from high-symmetry zone axes [001], [011], and [111], was that Bragg disks related by Friedel symmetry ($\boldsymbol{g}=(hkl)$ and $\boldsymbol{-g}=(\bar{h}\bar{k}\bar{l})$ tended to exhibit similar patterns of attention weights across most attention heads (Fig. S13--17; see Supplementary Movie S1). For these high-symmetry orientations, many reflections appear in symmetry-related positions of the reciprocal lattice. Among these positions, Bragg disks corresponding to $\boldsymbol{g}$ and $\boldsymbol{-g}$ (related by Friedel symmetry) share a similar local geometric configuration in reciprocal space, including identical radial distances $\left| \boldsymbol{g} \right|$ and comparable coordination environments. They also often exhibit similar reflection intensities for specific zone-axis directions. The observed similarity in attention weights for Friedel-related Bragg disks suggests that the lowest encoder layer may place emphasis on local geometric configuration and reflection intensities for orientation prediction.

Another notable feature appears in the attention patterns of certain attention heads in the lowest encoder layer for specific zone axes (Fig. S18). For diffraction patterns taken along the [001] zone axis, two attention heads in the lowest layer showed a striped pattern across many query indices. For these heads, each tended to place stronger attention on four key indices, corresponding to Bragg disks arranged with an approximate four-fold rotational symmetry in the diffraction pattern. A similar pattern appeared for diffraction patterns taken along the [111] zone axis. Two attention heads in the lowest layer again showed a striped structure across many queries, with each head placing noticeable attention on six key indices corresponding to Bragg disks arranged with an approximate pseudo-sixfold symmetry in the pattern. These findings suggest that certain heads in the lowest encoder layer may sometimes attend to subsets of Bragg disks related by (pseudo-)rotational symmetry in diffraction patterns, which may be relevant for orientation prediction.

However, these striped attention patterns were not consistently reproduced across independently trained models. When the model was trained five times independently using the same architecture and dataset, the corresponding attention patterns in the lowest encoder layer sometimes did not show the same striped feature or the same weighting of key indices related by rotational symmetry (Fig. S19,20). This lack of reproducibility indicates that these attention patterns should not be interpreted as a general description of how the model attends to Bragg disks. The striped attention patterns appear only in some models, for some diffraction patterns, and for some attention heads in the lowest encoder layer. Although these patterns are observed only in a subset of models and inputs, they do not provide reliable evidence that the transformer systematically encodes rotational symmetry in general. These observations are better interpreted as examples of model-specific behaviors that can arise during training rather than as generalizable or interpretable signatures of how the network processes symmetry in diffraction data.

\clearpage

\section*{Supplementary Text S2: Orientation sampling and simulation for generating training and validation diffraction datasets}

To generate the training and validation datasets, we adopted the following procedure.

We sampled symmetrically unique zone-axis directions $[UVW]$ on the surface of the unit sphere bounded by the $[001]$, $[011]$, and $[111]$ directions with an angular step size of 2 [$^\circ$], as shown in Fig. 2. For each zone-axis direction, thicknesses were sampled uniformly in the range $[3\sigma,900\sigma)$, where $\sigma$ denotes the lattice constant of the cubic crystals studied in this work. At this stage, diffraction patterns were simulated over the full thickness range solely to evaluate projected symmetry, and in-plane rotation angles were not yet sampled.

The simulated diffraction patterns were then examined to determine whether they exhibit mirror, rotational, or pseudo-rotational symmetry in the projected two-dimensional diffraction space (Fig. S12). When rotational or pseudo-rotational symmetry was present, the upper bound of the in-plane rotation angle was reduced accordingly, and in-plane angles were sampled with a step size of 1 [$\degree$]. For zone-axis directions whose diffraction patterns exhibit no mirror symmetry, the opposite zone-axis directions $[\bar{U}\bar{V}\bar{W}]$ were additionally sampled, since these orientations produce distinguishable diffraction patterns.

A total of eighty thickness values were subsequently selected by counting the number of Bragg disks in the diffraction patterns corresponding to each thickness. The thicknesses were sorted by Bragg-disk count. From this ordered list, 40 thicknesses with smaller numbers of Bragg disks were selected in ascending order, and 40 thicknesses with larger numbers of Bragg disks were selected in descending order. Diffraction patterns corresponding to intermediate thicknesses were not explicitly included, as their characteristic features were assumed to be effectively represented by data augmentations applied to the selected patterns (see Fig. S2 and \textit{Supplementary Text S3}).

Dynamical diffraction patterns \cite{zeltmann2023uncovering} were then simulated using the selected thicknesses, zone-axis directions, and in-plane rotation angles. For each simulated pattern, the token corresponding to the direct beam was removed, and the intensities of the remaining Bragg disks were normalized by the maximum Bragg disk intensity. Each crystal orientation, defined by the zone-axis direction and in-plane rotation angle, together with its corresponding diffraction pattern, served as a label-input pair for model training.

\section*{Supplementary Text S3: Data augmentation of Bragg disk features}

Data augmentation includes random displacement of radial distance, polar angle, Cartesian coordinates, and intensity values, all drawn from Gaussian noise distributions, as well as multiplication of intensity values by Gaussian noise. In addition, a fraction of Bragg disks with weak intensities are randomly removed, and, with a small probability, all Bragg disks in a pattern are assigned the same intensity value. These operations are designed to mimic signal variations observed in experimental 4D-STEM data, such as those arising from imaging conditions, shape factors, strain, substrate scattering, or imperfect Bragg disk detection. The augmentations are applied on the fly during training so that the model learns to rely on the presence and geometric relationships of Bragg disks, rather than on signal variations. This procedure also serves as a form of regularization that helps suppress overfitting. After augmentation, the intensities of all Bragg disks are normalized by the maximum Bragg disk intensity to ensure a consistent intensity scale for the model input. All data augmentations are implemented using the torchvision software \cite{marcel2010torchvision}. Fig. S2 provides a schematic illustration of the data augmentation procedures used in this work.

\section*{Supplementary Text S4: A potential route for utilizing model predictions on noisy experimental 4D-STEM data}

Within the highly correlated domains, the model predictions closely match those obtained from template matching \cite{cautaerts2022free,ophus2022automated,rauch2010automated,rauch2014automated}. This close agreement suggests a potential approach for utilizing model predictions on noisy experimental 4D-STEM data without ground-truth labels. In this approach, one may first apply the model to rapidly obtain orientation estimates across the scan grid. Next, regions where the predicted zone-axis directions are spatially coherent across neighboring scan positions can be identified and defined as highly correlated domains. Within these domains, the spatial coherence of the model predictions indicates that the structural features in the diffraction patterns are broadly similar across the scan grid. Under these conditions, the model may leverage these features to produce reasonable orientation estimates comparable to those from template matching. Finally, for the remaining regions, where the model predictions exhibit weaker spatial coherence, template matching may be applied to obtain more precise orientation assignments.

We note that this proposed route is derived from observations on a single experimental dataset. The extent to which it applies to other materials systems, experimental conditions, and noise regimes remains to be evaluated.

\section*{Supplementary Text S5: Training configuration and loss convergence}

Model optimization was performed using the AdamW optimizer \cite{loshchilov2017decoupled}. The learning rate was linearly warmed up to $7\times10^{-5}$ over the first 15 epochs and subsequently annealed using a cosine decay schedule to a minimum value of $5\times10^{-7}$. Training was conducted for 265 epochs with a mini-batch size of $1024$. 

For orientation prediction, we performed six independent training runs. For the joint prediction of orientations and phases, we performed three independent training runs. In several runs, the training and validation losses exhibited an extended plateau before decreasing further at later epochs (Fig. 6b). This behavior indicates sensitivity to initialization and optimization dynamics rather than a systematic instability of the model. Future work may explore strategies such as self-supervised pretraining or adaptive learning-rate schedules to improve optimization dynamics.


\section*{Supplementary Figures S1-S20}

\begin{figure}[htb]
\begin{center}
\includegraphics[width=0.99\textwidth]{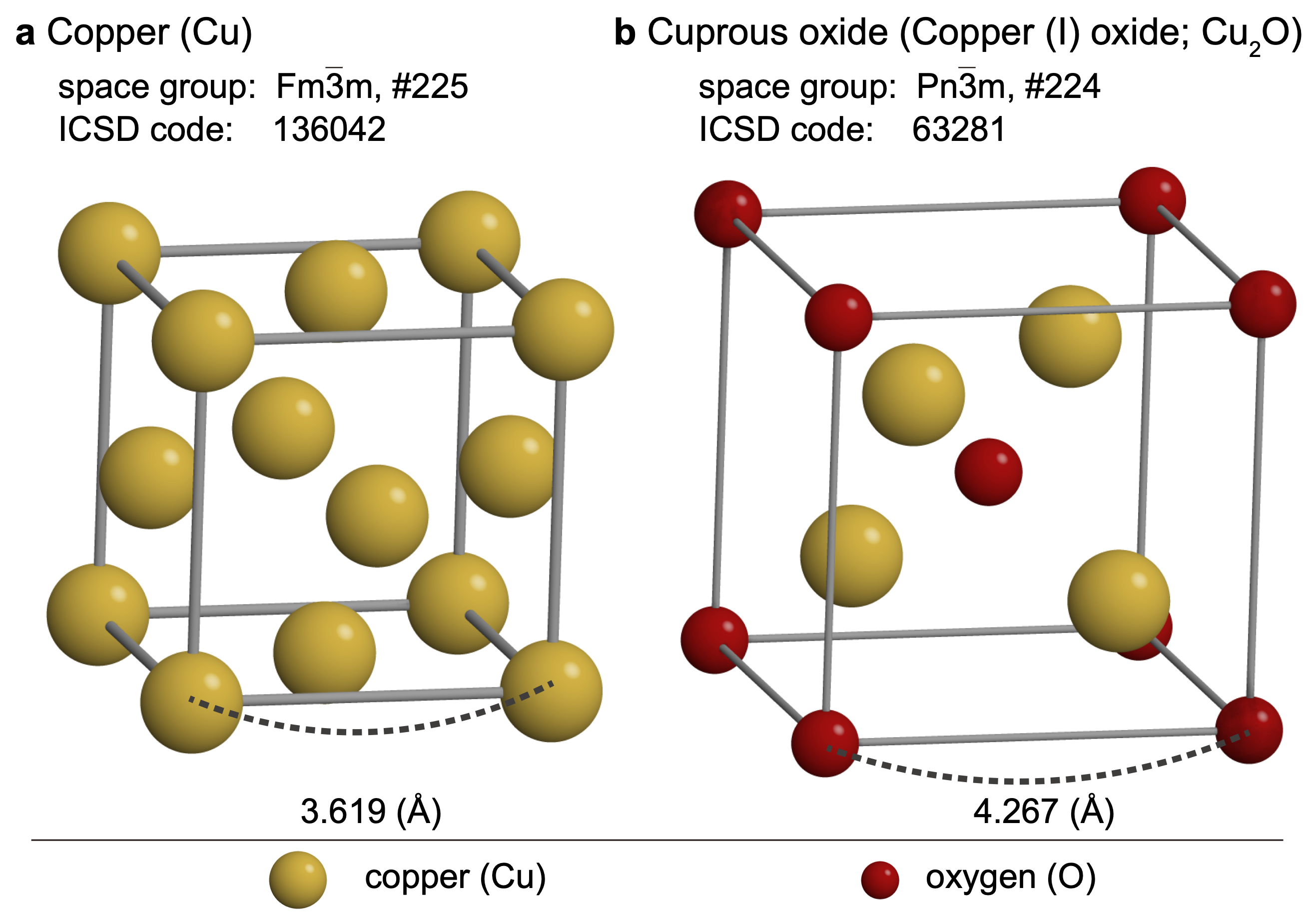}
\vspace*{-3mm}
\caption{\textbf{a},\textbf{b} Schematic illustrations of a face-centered-cubic (fcc) unit cell of copper ($Cu$; space group $Fm\bar{3}m$) (a) and a cubic unit cell of Cuprous oxide ($Cu_{2}O$; space group $Pn\bar{3}m$). The unit cells are retrieved from the Inorganic Crystal Structure Database (ICSD) \cite{belsky2002new}.}
\label{fig:FIG_S1}
\end{center}
\end{figure}

\begin{figure}[htb]
\begin{center}
\includegraphics[width=0.99\textwidth]{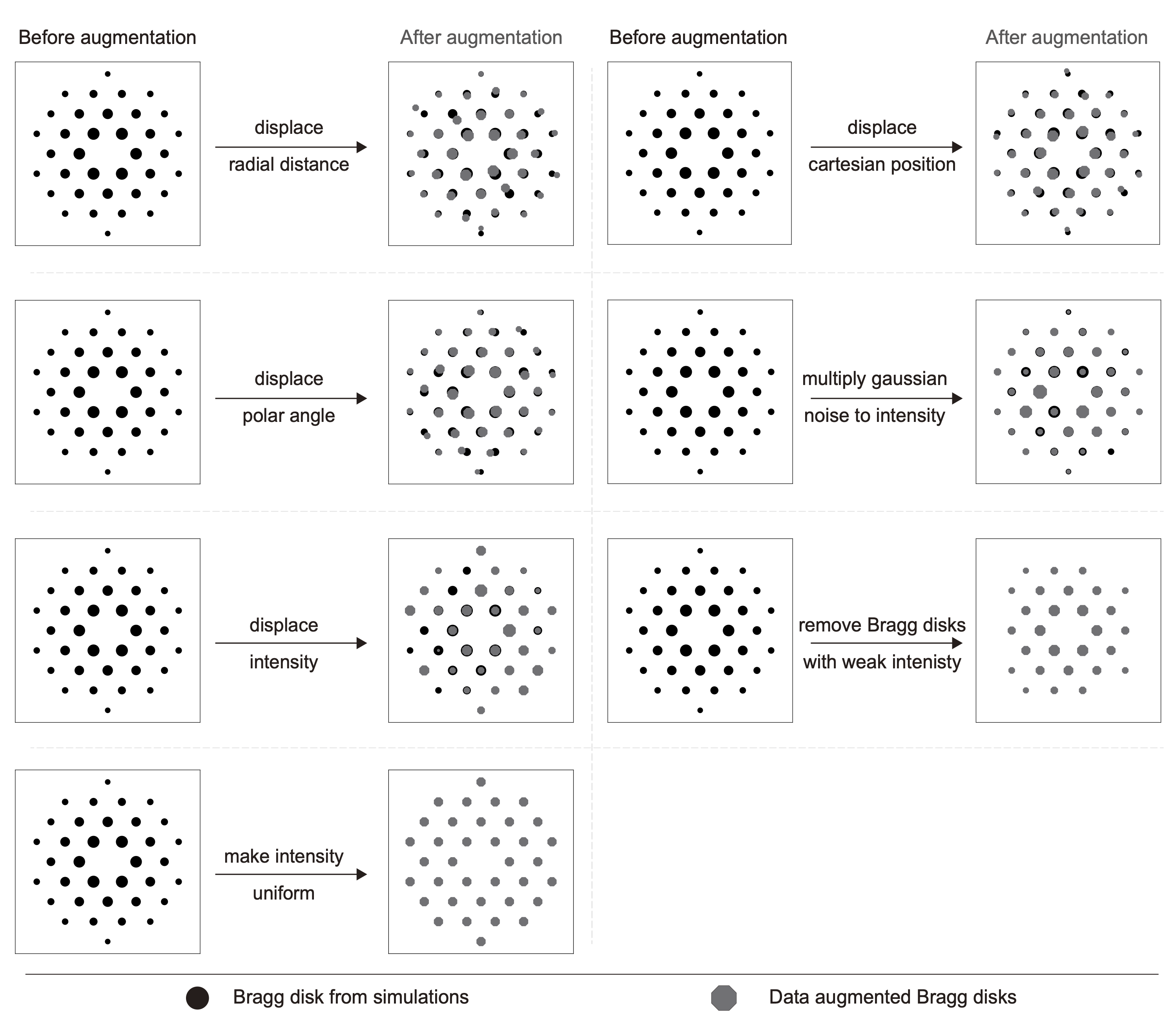}
\vspace*{-3mm}
\caption{Schematic illustration of data augmentations used in this work. Black circles and gray octagons represent simulated Bragg disks and data-augmented Bragg disks, respectively. The size of the circles and octagons represents the intensities of the Bragg disks.}
\label{fig:FIG_S2}
\end{center}
\end{figure}

\begin{figure}[htb]
\begin{center}
\includegraphics[width=0.99\textwidth]{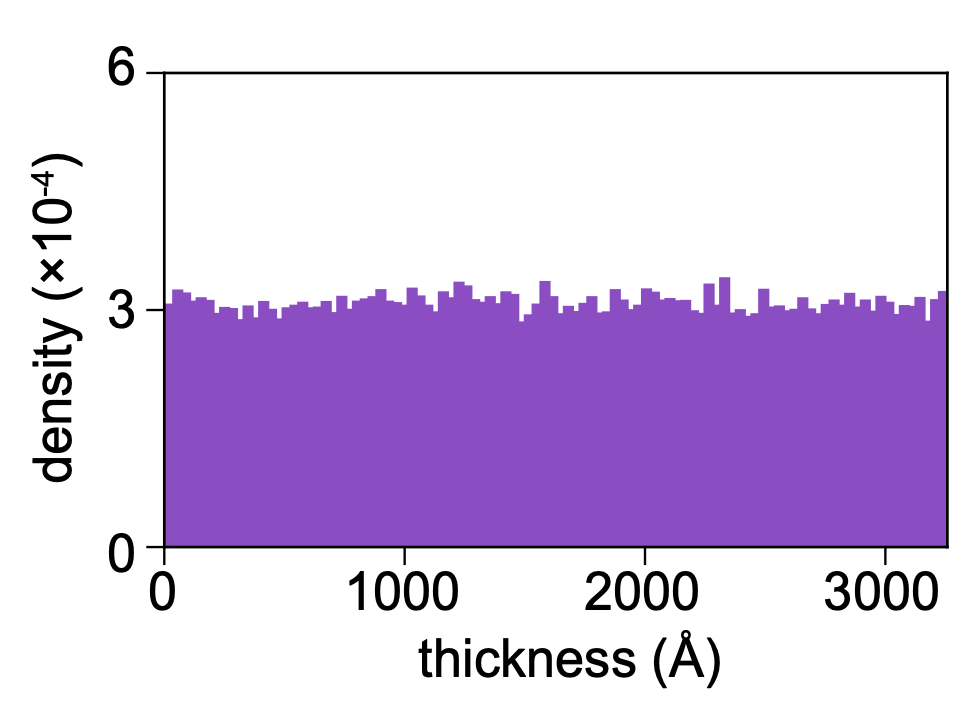}
\vspace*{-3mm}
\caption{The thickness values uniformly sampled for generating the synthetic test set.}
\label{fig:FIG_S3}
\end{center}
\end{figure}

\begin{figure}[htb]
\begin{center}
\includegraphics[width=0.95\textwidth]{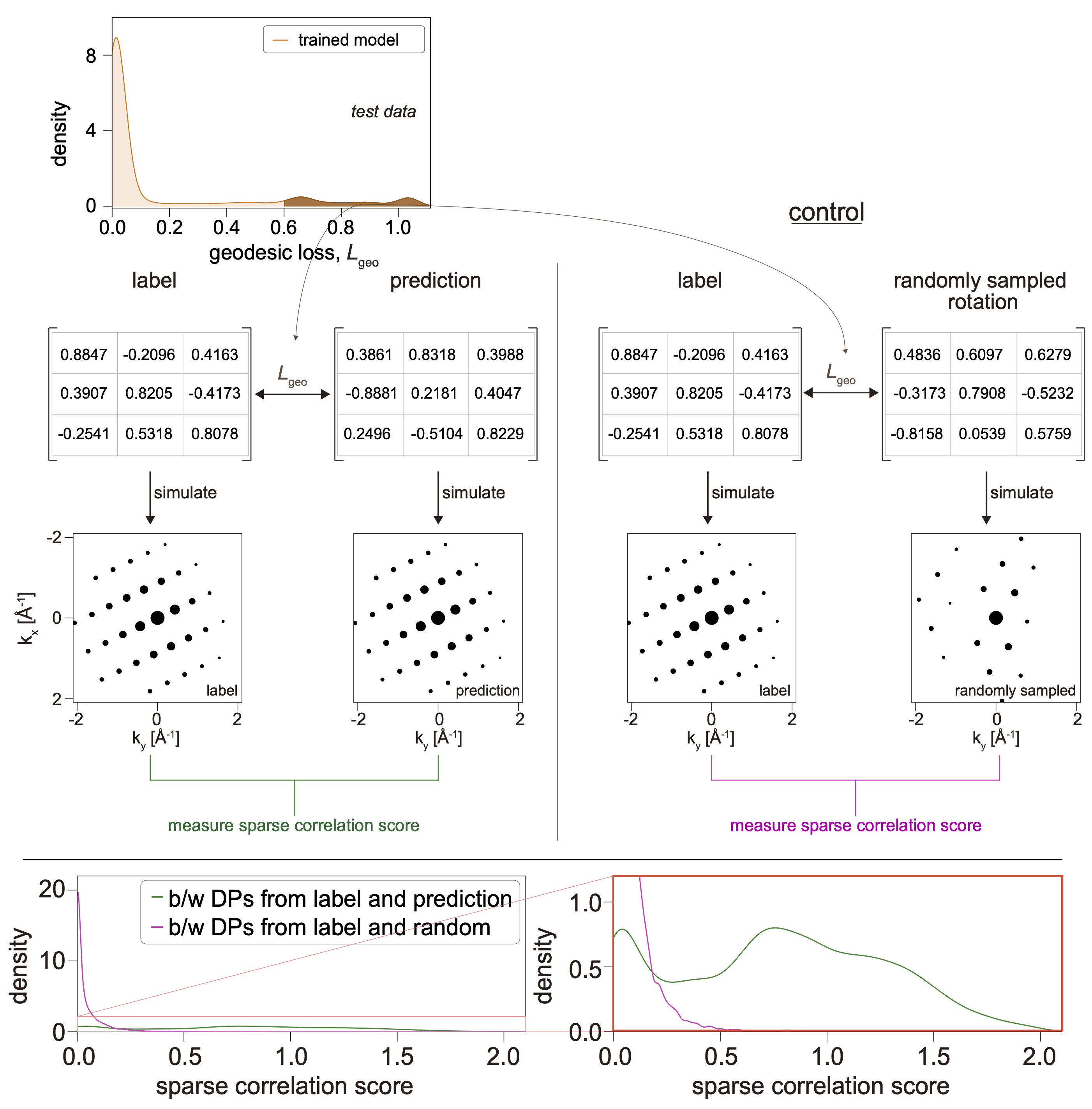}
\vspace*{-3mm}
\caption{Sparse correlation scores for test-set model predictions with large geodesic loss. Data points with geodesic loss greater than 0.6 between the model prediction and the label orientation are selected. For these data points, sparse correlation scores are calculated between the simulated Bragg disks from the model prediction and from the label orientation (top left). For reference, for each label orientation, a random orientation yielding the same geodesic loss is sampled, and the sparse correlation score between the simulated Bragg disks from the label orientation and from the randomly sampled orientation is calculated (top right). Although the model predictions and the randomly sampled orientations have identical geodesic loss values for each label, the distribution of sparse correlation scores for the randomly sampled orientations shows a larger population at low correlation scores (down). In contrast, the model predictions tend to yield higher sparse correlation scores, indicating closer agreement between the corresponding diffraction patterns.}
\label{fig:FIG_S4}
\end{center}
\end{figure}

\begin{figure}[htb]
\begin{center}
\includegraphics[width=0.99\textwidth]{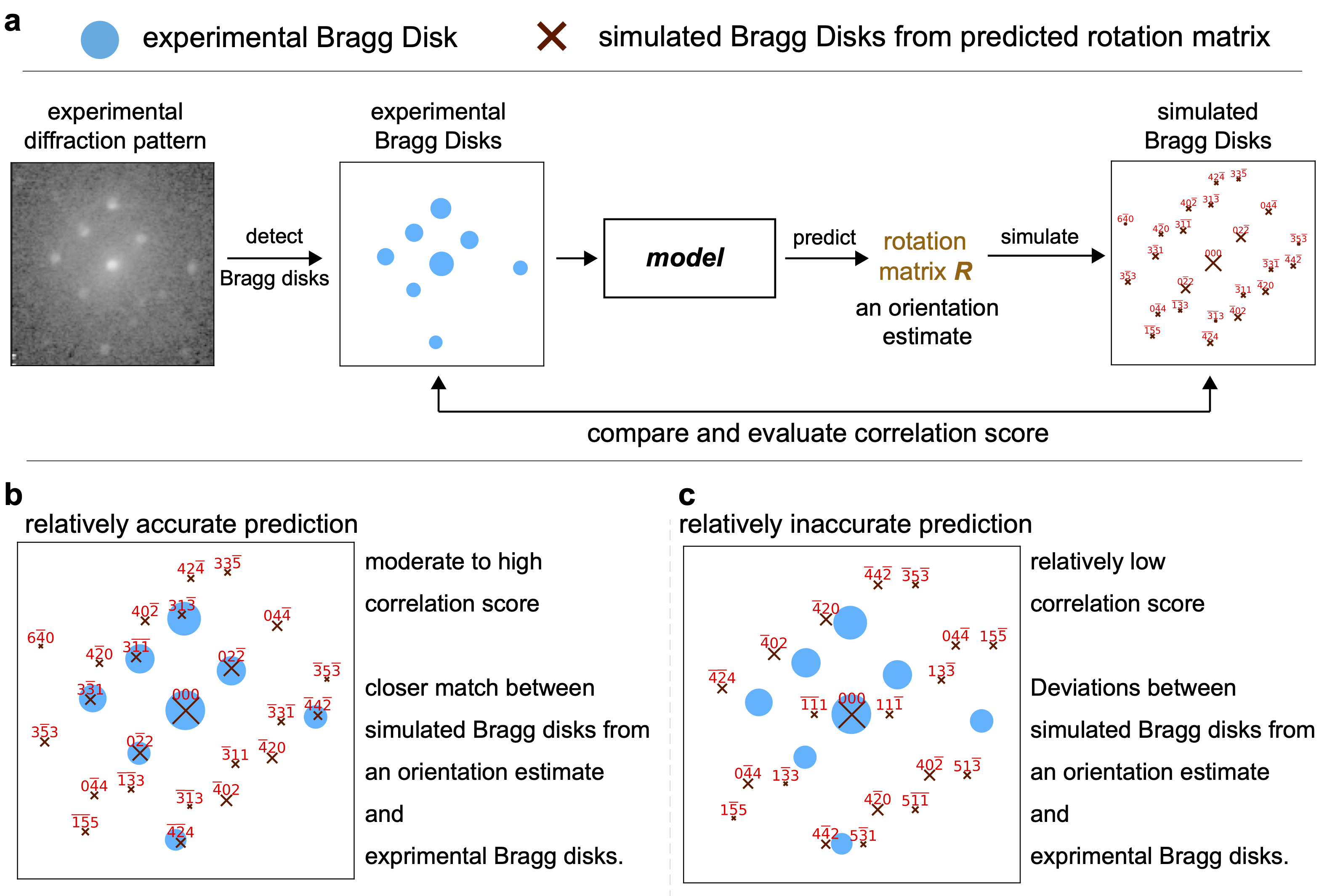}
\vspace*{-3mm}
\caption{\textbf{a} Protocol for evaluating the accuracy of orientation estimates. The correlation score is calculated between simulated Bragg disks from an orientation estimate and the corresponding experimental data. \textbf{b,c} Schematic illustrations of relatively accurate (b) and inaccurate (c) orientation estimates, yielding high and low correlation scores, respectively. Red integers overlaid on each simulated Bragg disk indicate the corresponding Miller indices.}
\label{fig:FIG_S5}
\end{center}
\end{figure}

\vspace{-0.5cm} 

\begin{figure}[htb]
\begin{center}
\includegraphics[width=0.99\textwidth]{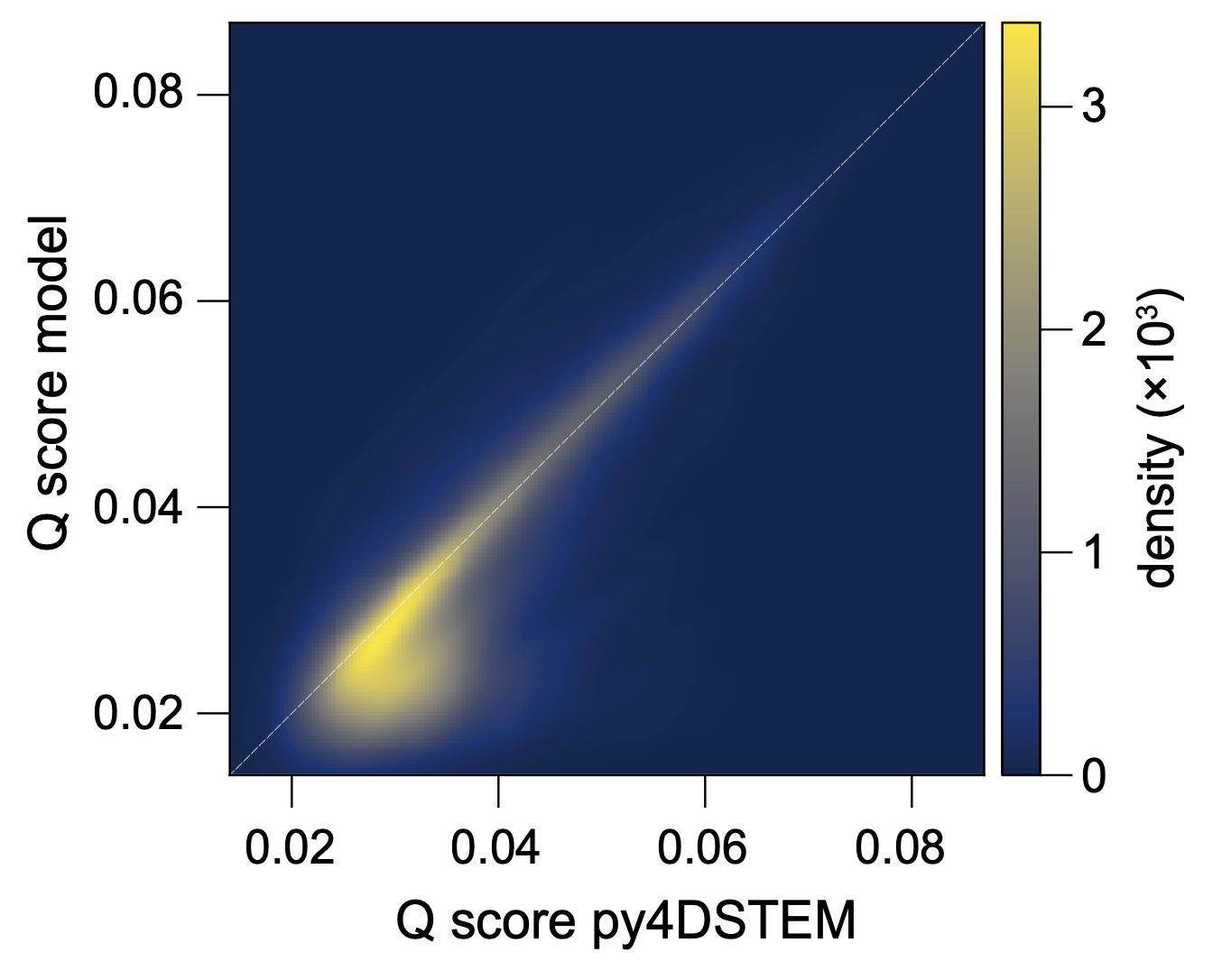}
\vspace*{-3mm}
\caption{Two-dimensional density map showing the joint distribution of Q scores obtained from model predictions (y-axis) and py4DSTEM template matching (x-axis). Color intensity represents the density of observations.}
\label{fig:FIG_S6}
\end{center}
\end{figure}

\begin{figure}[htb]
\begin{center}
\includegraphics[width=0.99\textwidth]{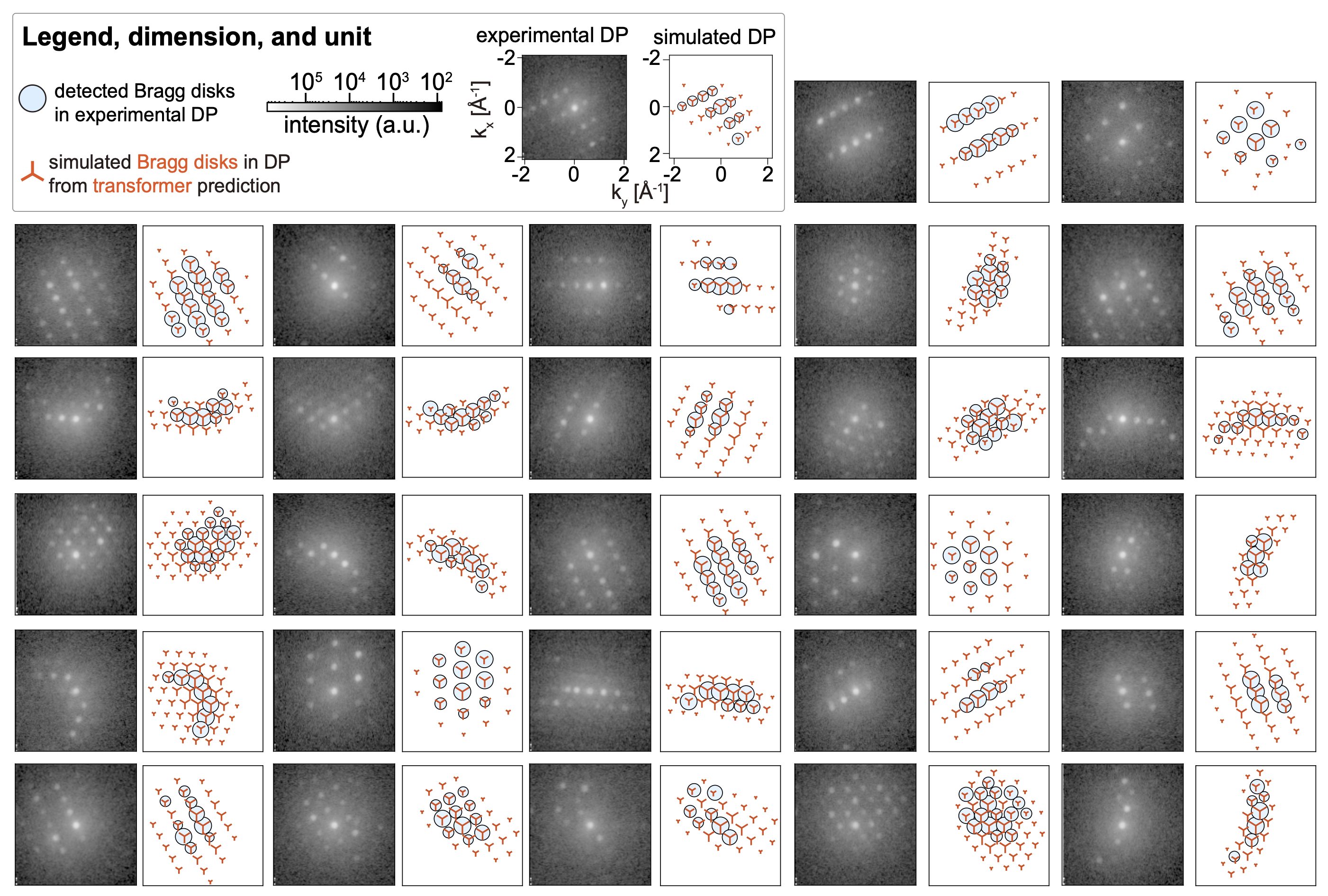}
\vspace*{-3mm}
\caption{Examples of model predictions yielding close agreement between simulated and experimental diffraction patterns. For each example, the experimental diffraction pattern (left) and the simulated diffraction pattern (right) are shown. To illustrate the correspondence between simulated and experimental Bragg disks, the experimental Bragg disks (blue) are overlaid on the simulated diffraction patterns.}
\label{fig:FIG_S7}
\end{center}
\end{figure}

\begin{figure}[htb]
\begin{center}
\includegraphics[width=0.99\textwidth]{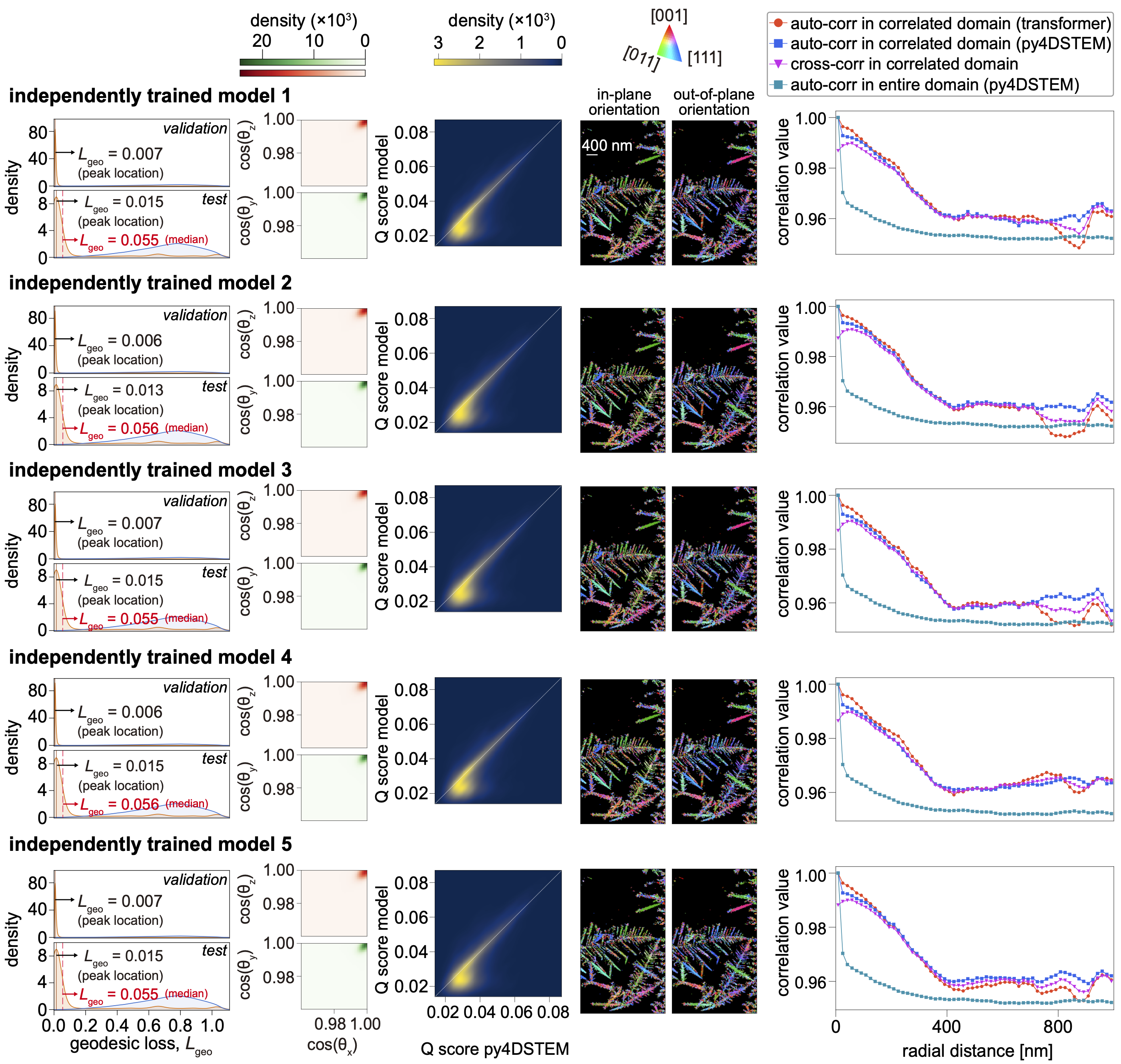}
\vspace*{-3mm}
\caption{Predictions from five independently trained models, in addition to those shown in the main figure. For each model, we show the densities of the geodesic loss for the validation and test datasets, the density of cosines of the angular misalignment between symmetry-reduced crystallographic axes, the density of $Q$ scores from transformer predictions and py4DSTEM template matching, the corresponding orientation map, and the radially averaged two-point correlation functions.}
\label{fig:FIG_S8}
\end{center}
\end{figure}

\begin{figure}[htb]
\begin{center}
\includegraphics[width=0.99\textwidth]{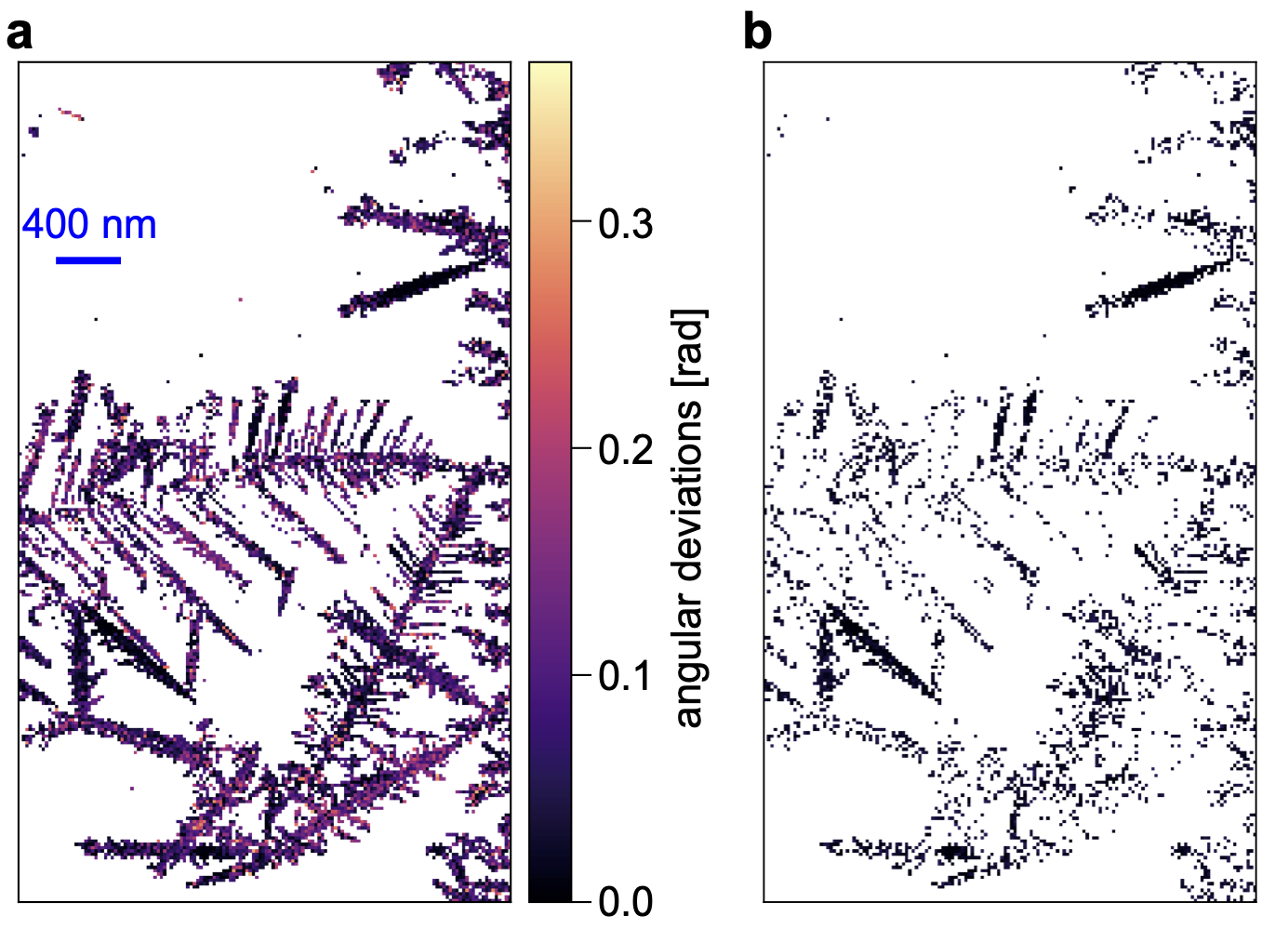}
\vspace*{-3mm}
\caption{\textbf{a} Directional uncertainty of the predicted zone-axis across the scan positions. For each pixel, the angular deviation between each model’s prediction and the mean across six models is shown. Pixels with no assigned orientations are masked in white, while colored pixels indicate angular deviations. \textbf{b} Regions of low directional uncertainty, with angular deviations below $0.052$ [rad] ($3$ [$^\circ$]), highlighting areas where model predictions are highly consistent.}
\label{fig:FIG_S9}
\end{center}
\end{figure}

\begin{figure}[htb]
\begin{center}
\includegraphics[width=0.99\textwidth]{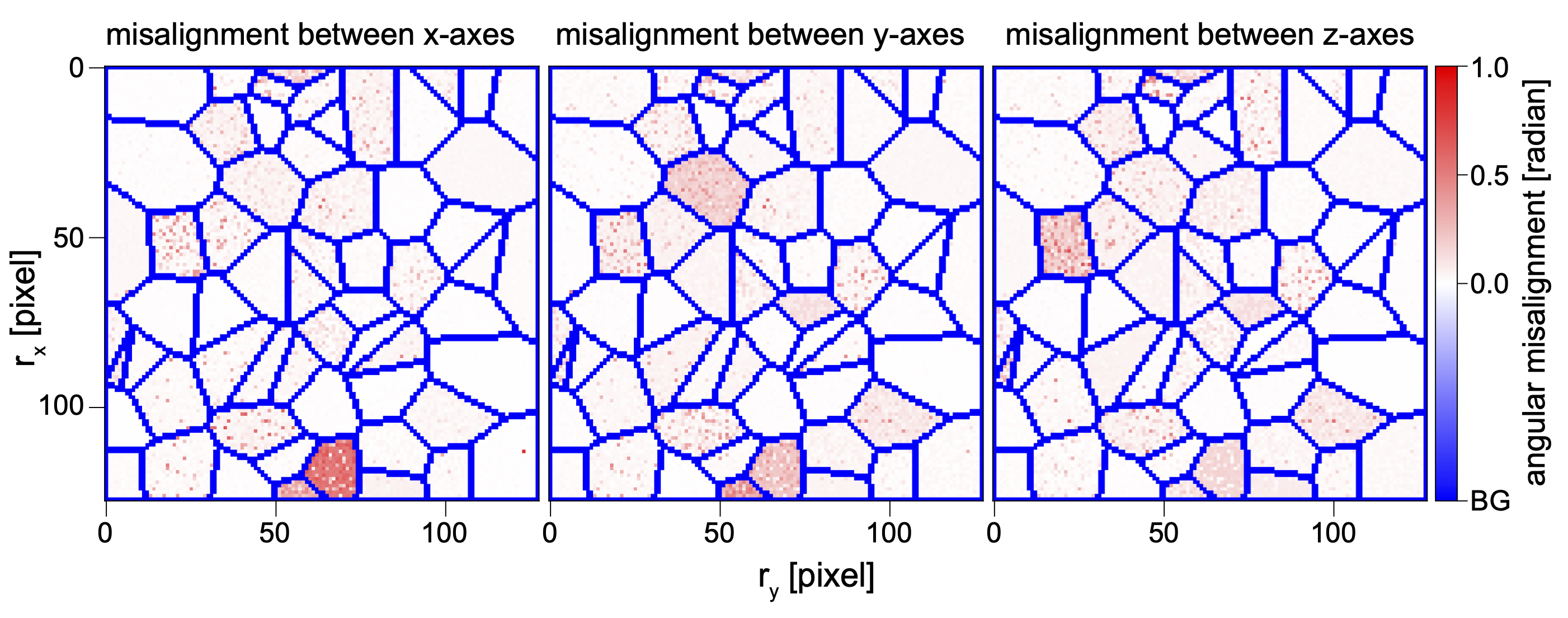}
\vspace*{-3mm}
\caption{Angular misalignment between symmetry-reduced crystallographic directions.
For each scan position in the synthetic 4D-STEM dataset shown in Fig. 6, the angular misalignments between the symmetry-reduced x (left), y (middle), and z (right) directions derived from the label orientation and those predicted by the model are shown.}
\label{fig:FIG_S10}
\end{center}
\end{figure}

\begin{figure}[htb]
\begin{center}
\includegraphics[width=0.99\textwidth]{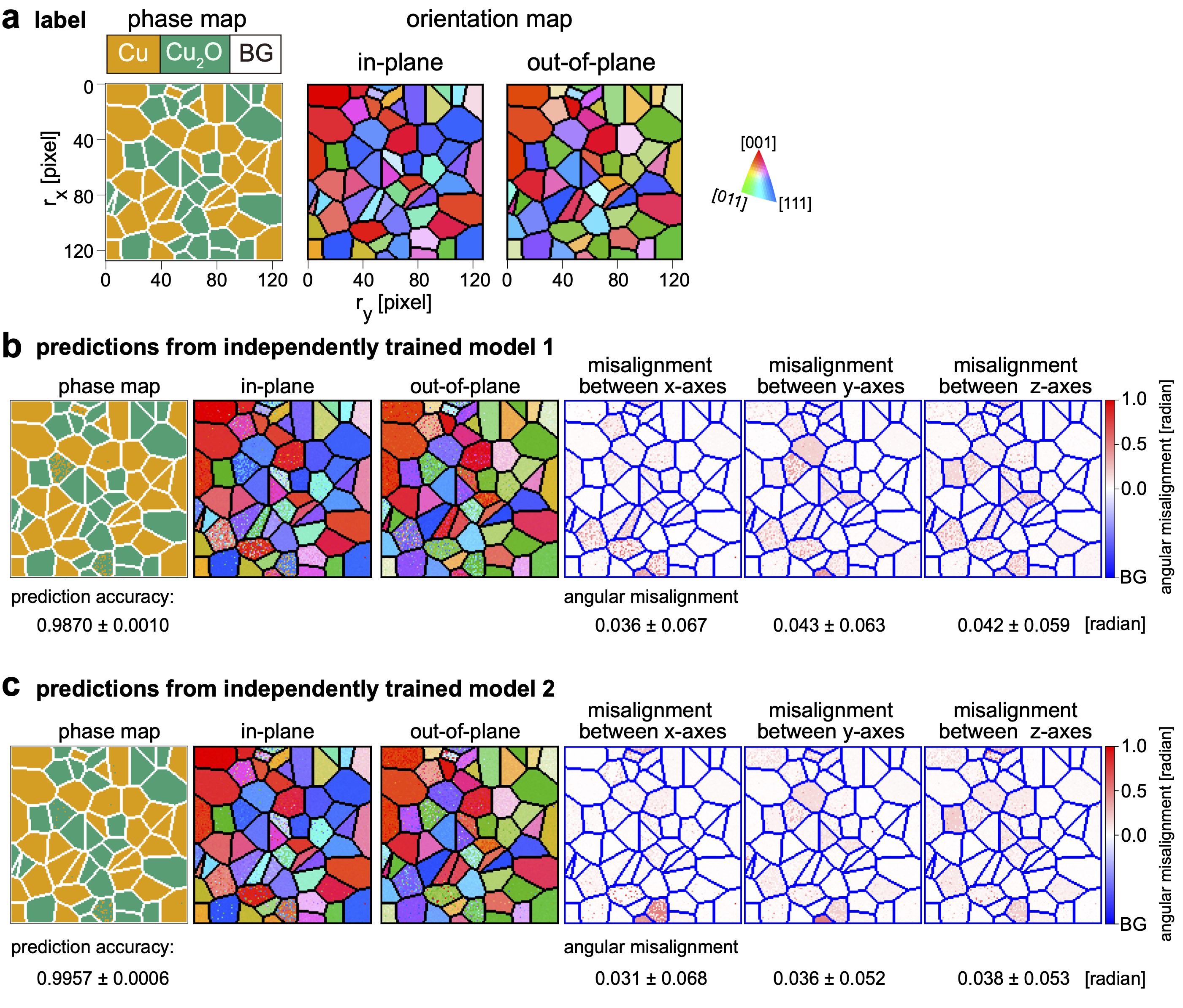}
\vspace*{-3mm}
\caption{Joint predictions from two independently trained models, in addition to those shown in Fig.~6. For each model, we show the phase map, orientation map, and angular misalignment between symmetry-reduced x-, y-, and z-axes.}
\label{fig:FIG_S11}
\end{center}
\end{figure}

\begin{figure}[htb]
\begin{center}
\includegraphics[width=0.99\textwidth]{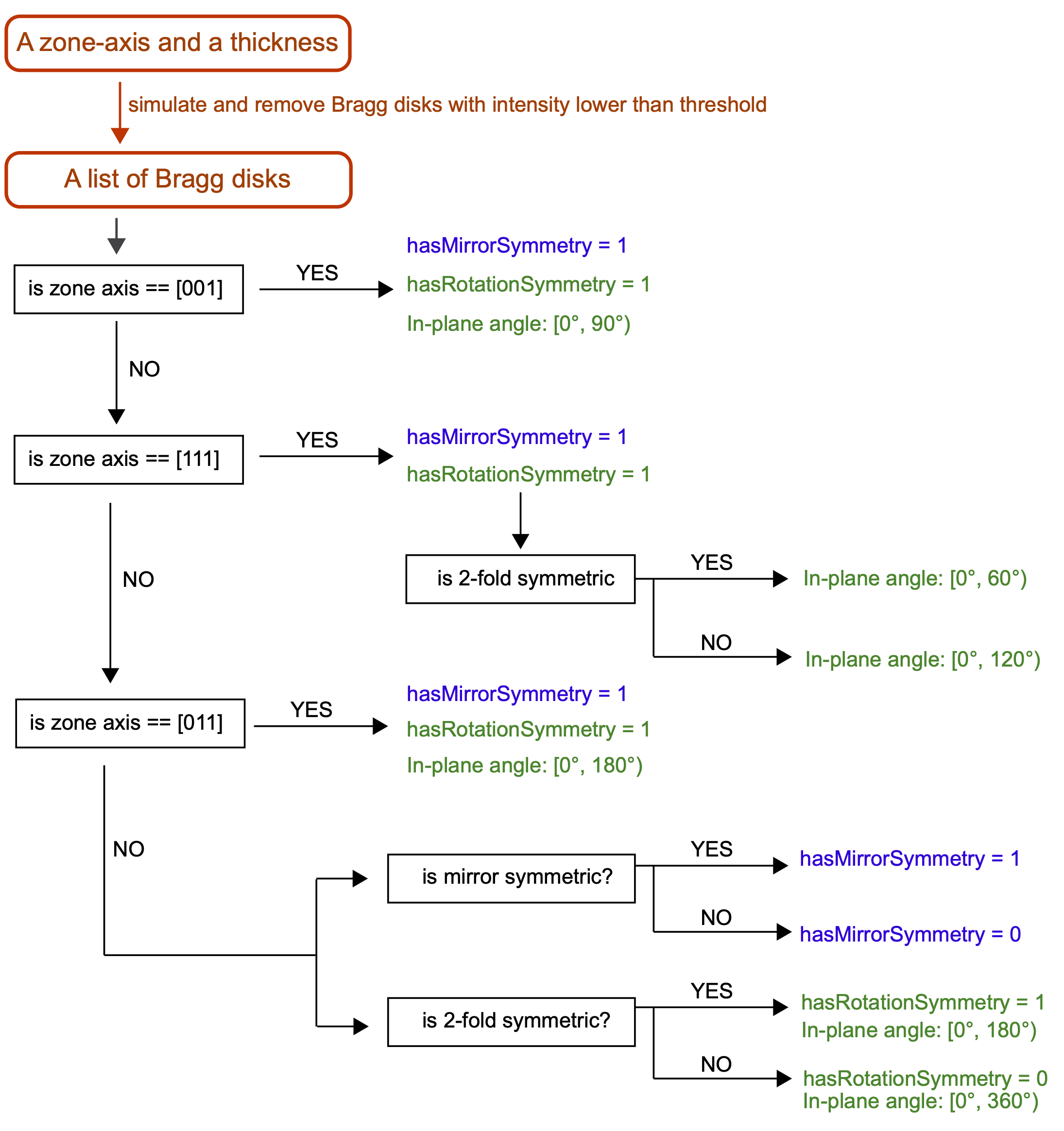}
\vspace*{-3mm}
\caption{Protocol for determining the upper bound of in-plane rotation and the presence of mirror symmetry for a sampled zone axis.}
\label{fig:FIG_S12}
\end{center}
\end{figure}

\begin{figure}[htb]
\begin{center}
\includegraphics[width=0.99\textwidth]{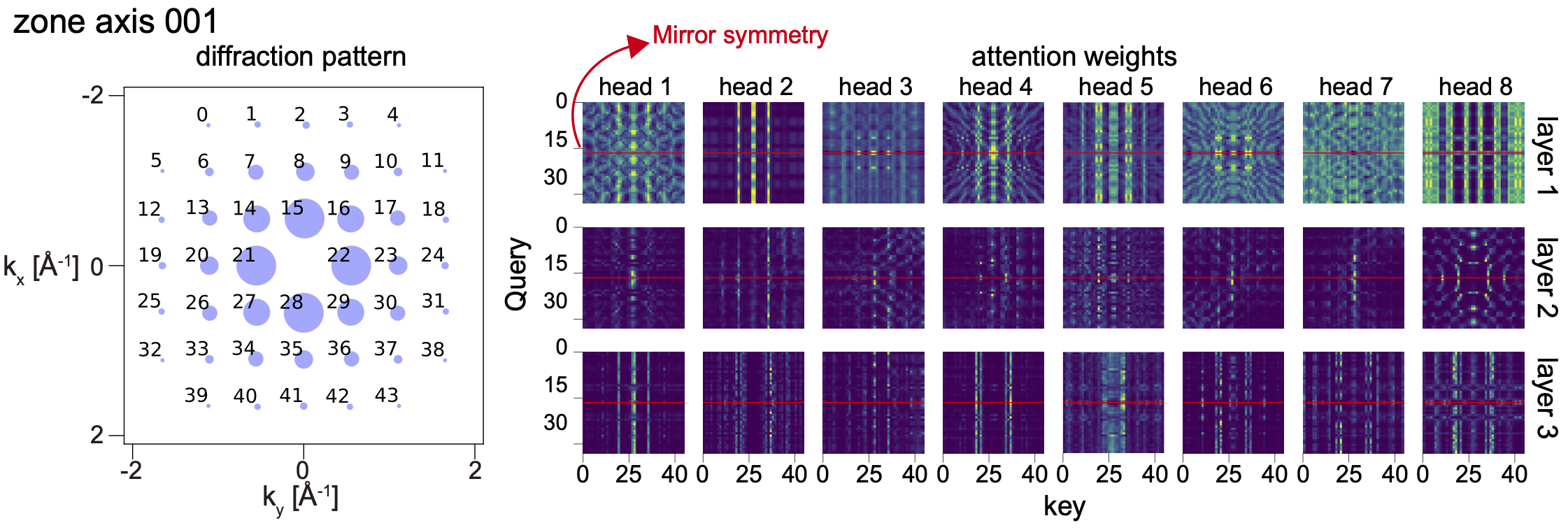}
\vspace*{-3mm}
\caption{(left) A diffraction pattern taken along the [001] zone axis. The number shown at the top left of each Bragg disk indicates its query index. (right) Attention weights from three encoder layers for the corresponding diffraction pattern. For the attention weights from the lowest encoder layer (layer 1), an approximate mirror symmetry is observed about the line at query-axis value 22. Because Bragg disks related by Friedel symmetry appear at mirrored positions along the query axis, this symmetry suggests similar attention weights for Friedel-related disk pairs. Such symmetry is less apparent in deeper encoder layers.}
\label{fig:FIG_S13}
\end{center}
\end{figure}

\begin{figure}[htb]
\begin{center}
\includegraphics[width=0.99\textwidth]{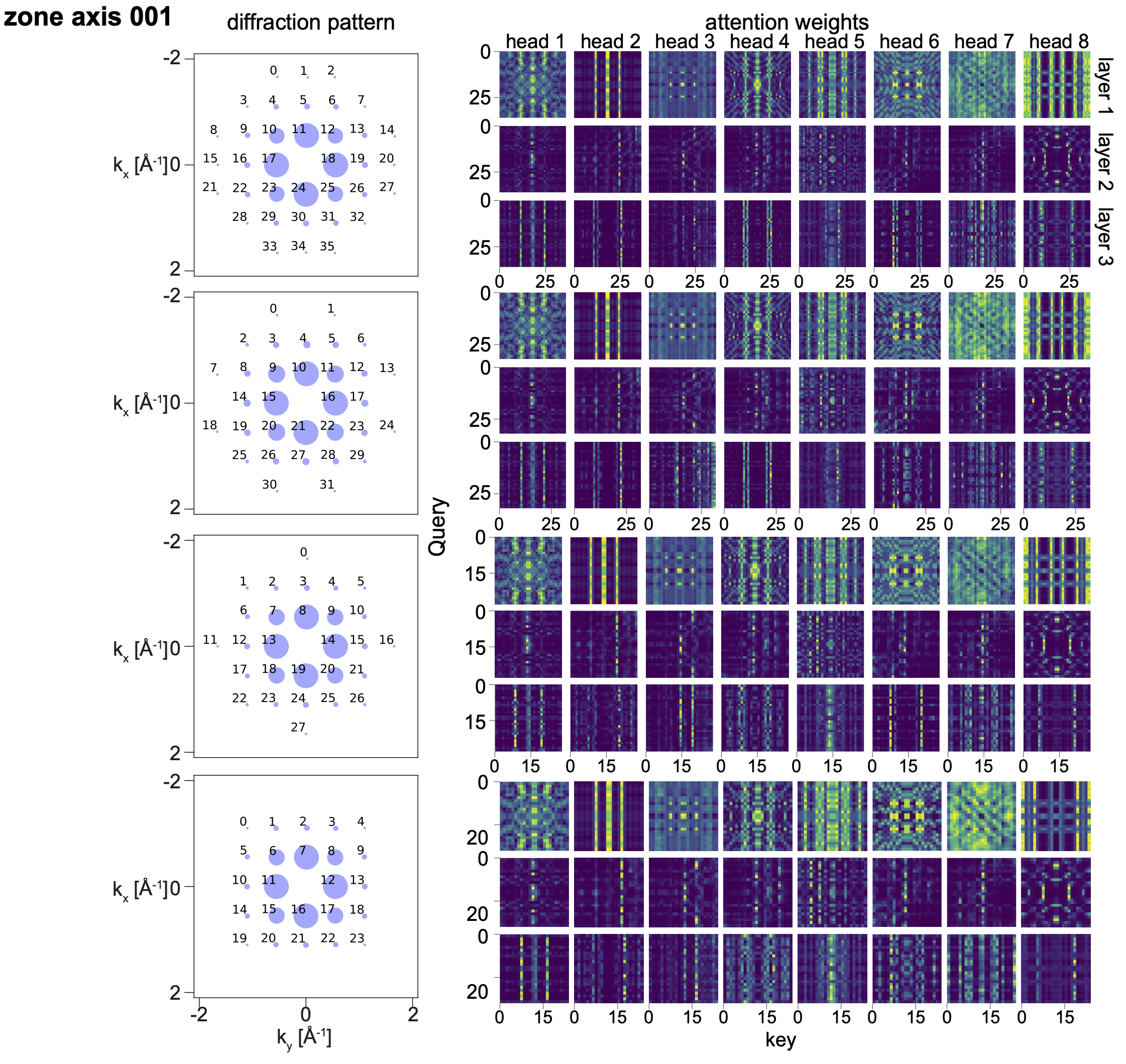}
\vspace*{-3mm}
\caption{(left) Diffraction patterns taken from four different specimen thicknesses along the [001] zone axis. (right) Attention weights from three encoder layers for the corresponding diffraction patterns. For the attention weights from the lowest encoder layer, Bragg disks related by Friedel symmetry show similar attention weights.}
\label{fig:FIG_S14}
\end{center}
\end{figure}

\begin{figure}[htb]
\begin{center}
\includegraphics[width=0.99\textwidth]{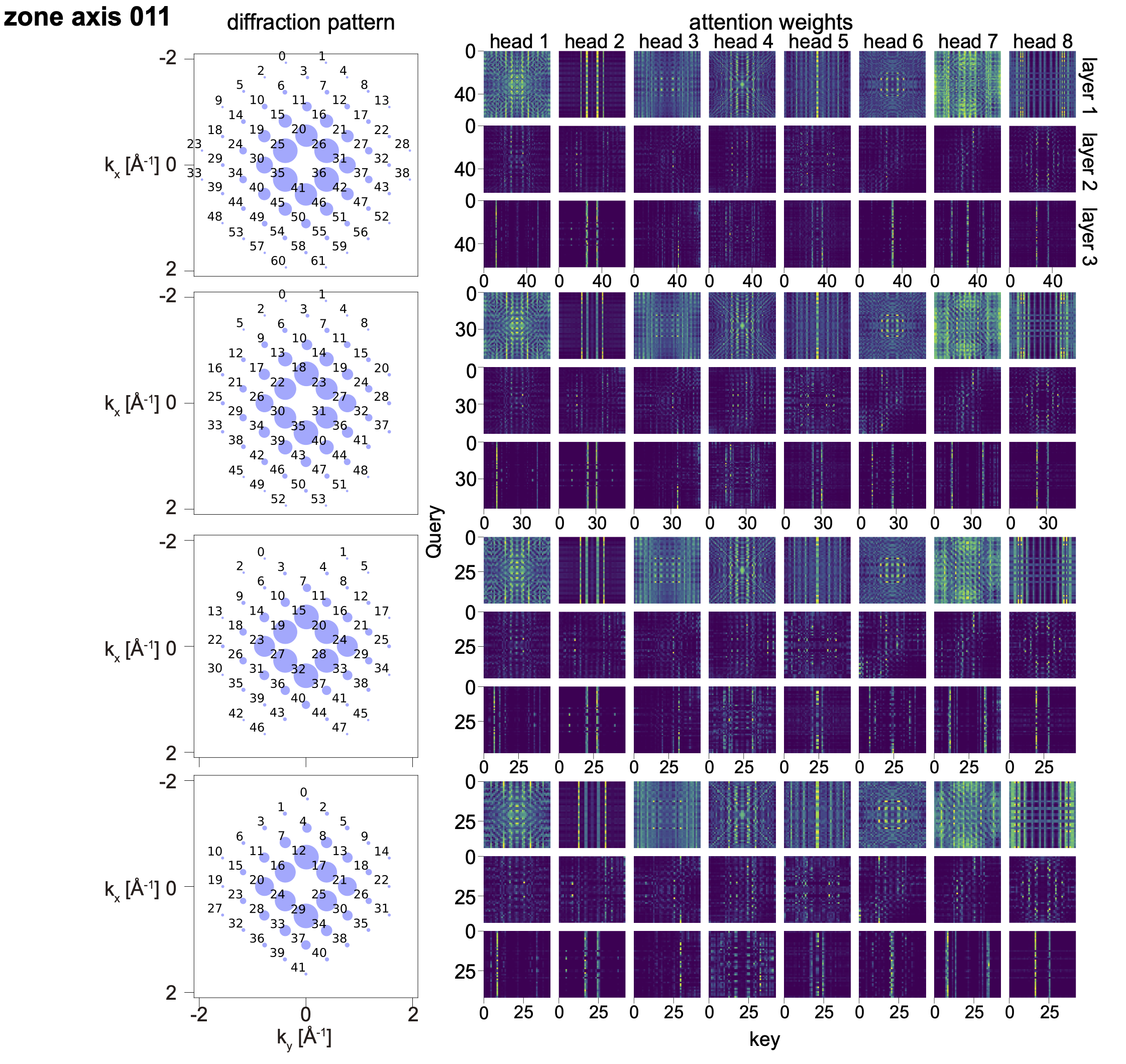}
\vspace*{-3mm}
\caption{(left) Diffraction patterns taken from four different specimen thicknesses along the [011] zone axis. (right) Attention weights from three encoder layers for the corresponding diffraction patterns.}
\label{fig:FIG_S15}
\end{center}
\end{figure}

\begin{figure}[htb]
\begin{center}
\includegraphics[width=0.99\textwidth]{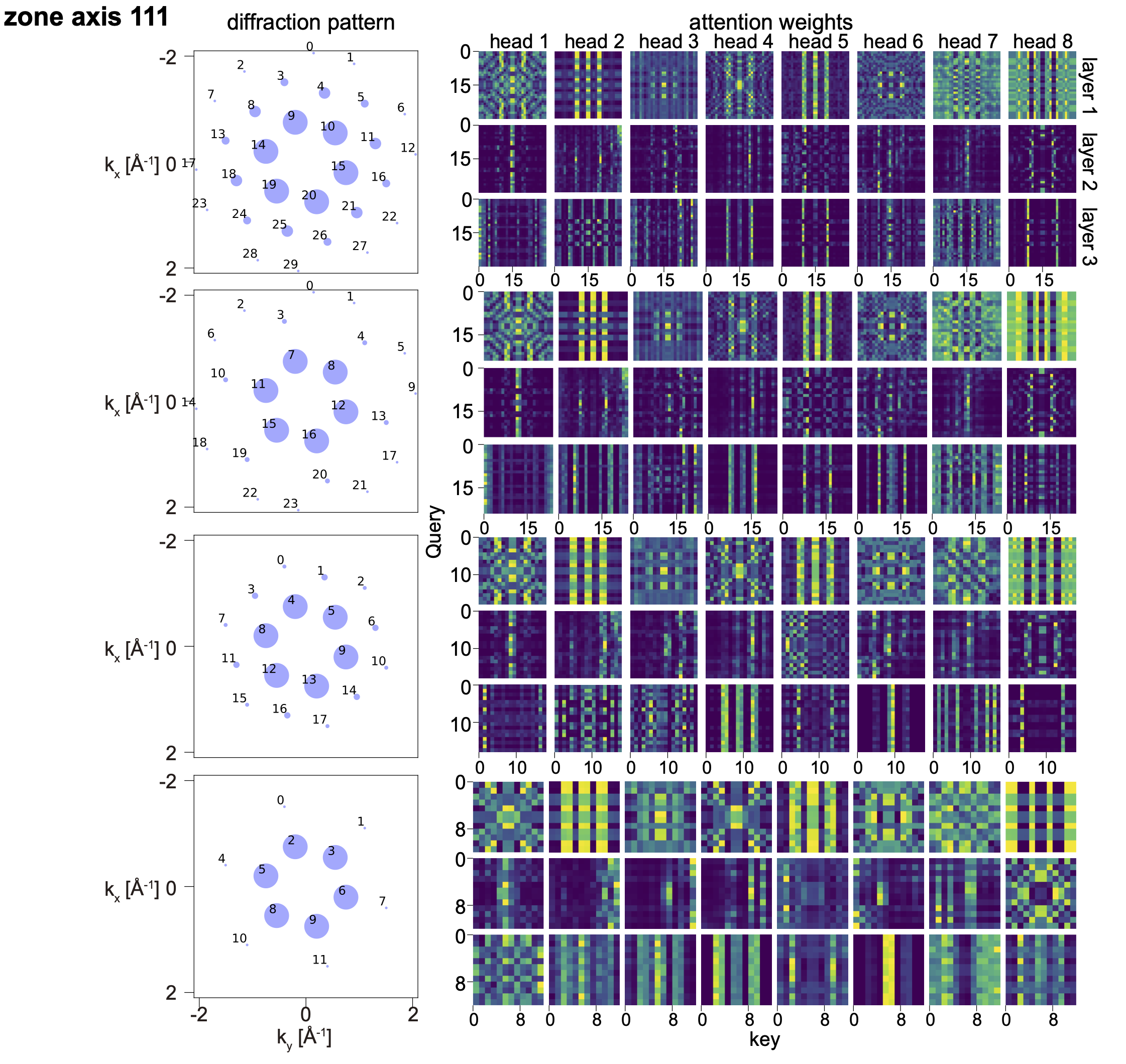}
\vspace*{-3mm}
\caption{(left) Diffraction patterns taken from four different specimen thicknesses along the [111] zone axis. (right) Attention weights from three encoder layers for the corresponding diffraction patterns.}
\label{fig:FIG_S16}
\end{center}
\end{figure}

\begin{figure}[htb]
\begin{center}
\includegraphics[width=0.99\textwidth]{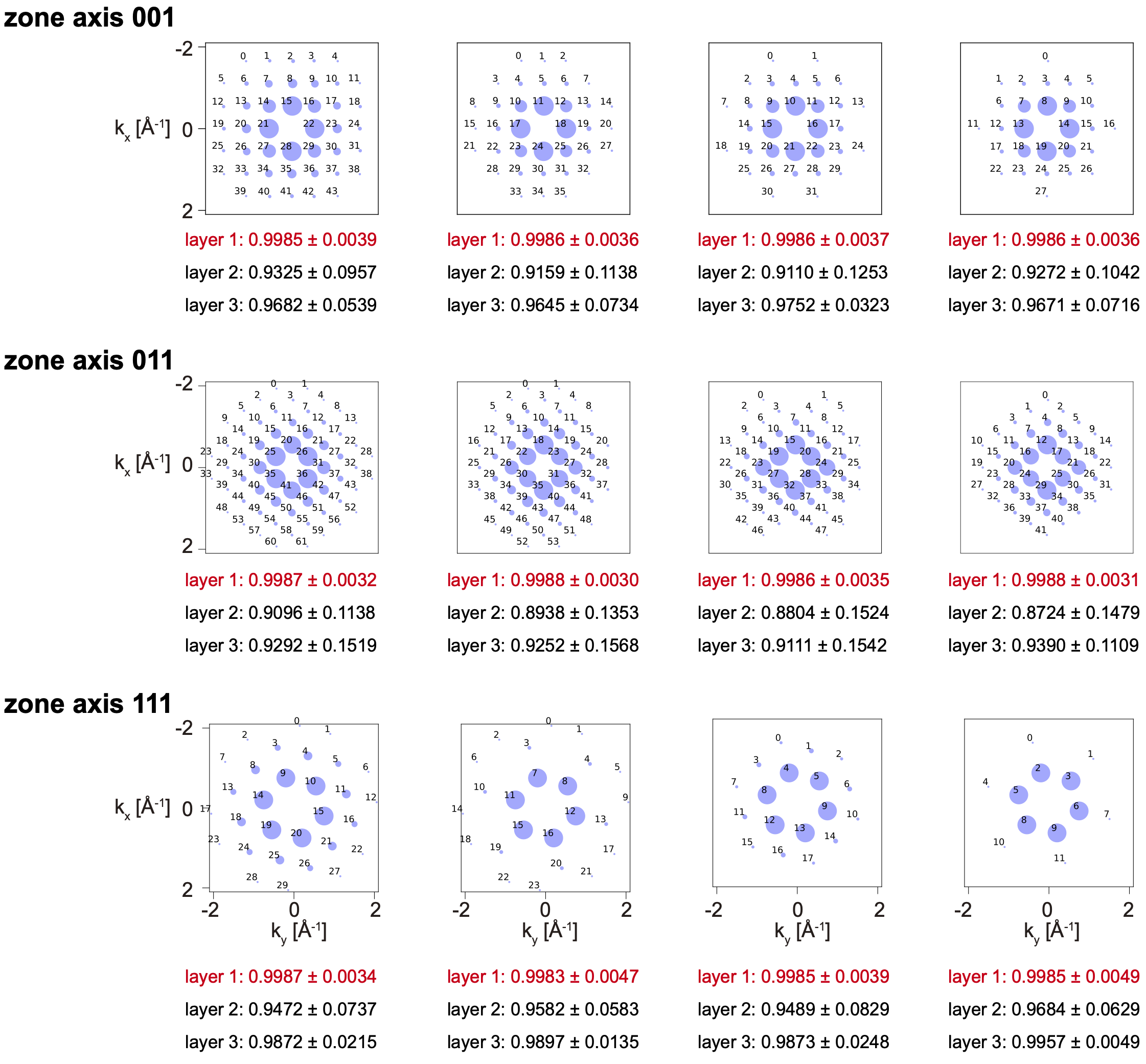}
\vspace*{-3mm}
\caption{Average cosine similarity of attention weights for Bragg disks related by Friedel symmetry in each encoder layer. For the shown diffraction patterns, attention weights were calculated for each encoder layer. For each attention head, the cosine similarity was computed between attention-weight vectors corresponding to Bragg disks related by Friedel symmetry. These similarities were first averaged over all Friedel-related disk pairs within each head and then averaged across heads to obtain the mean cosine similarity for that encoder layer. For all diffraction patterns, the average cosine similarity in the lowest encoder layer (red) is greater than 0.99, indicating that Bragg disks related by Friedel symmetry show highly similar attention weights.}
\label{fig:FIG_S17}
\end{center}
\end{figure}

\begin{figure}[htb]
\begin{center}
\includegraphics[width=0.99\textwidth]{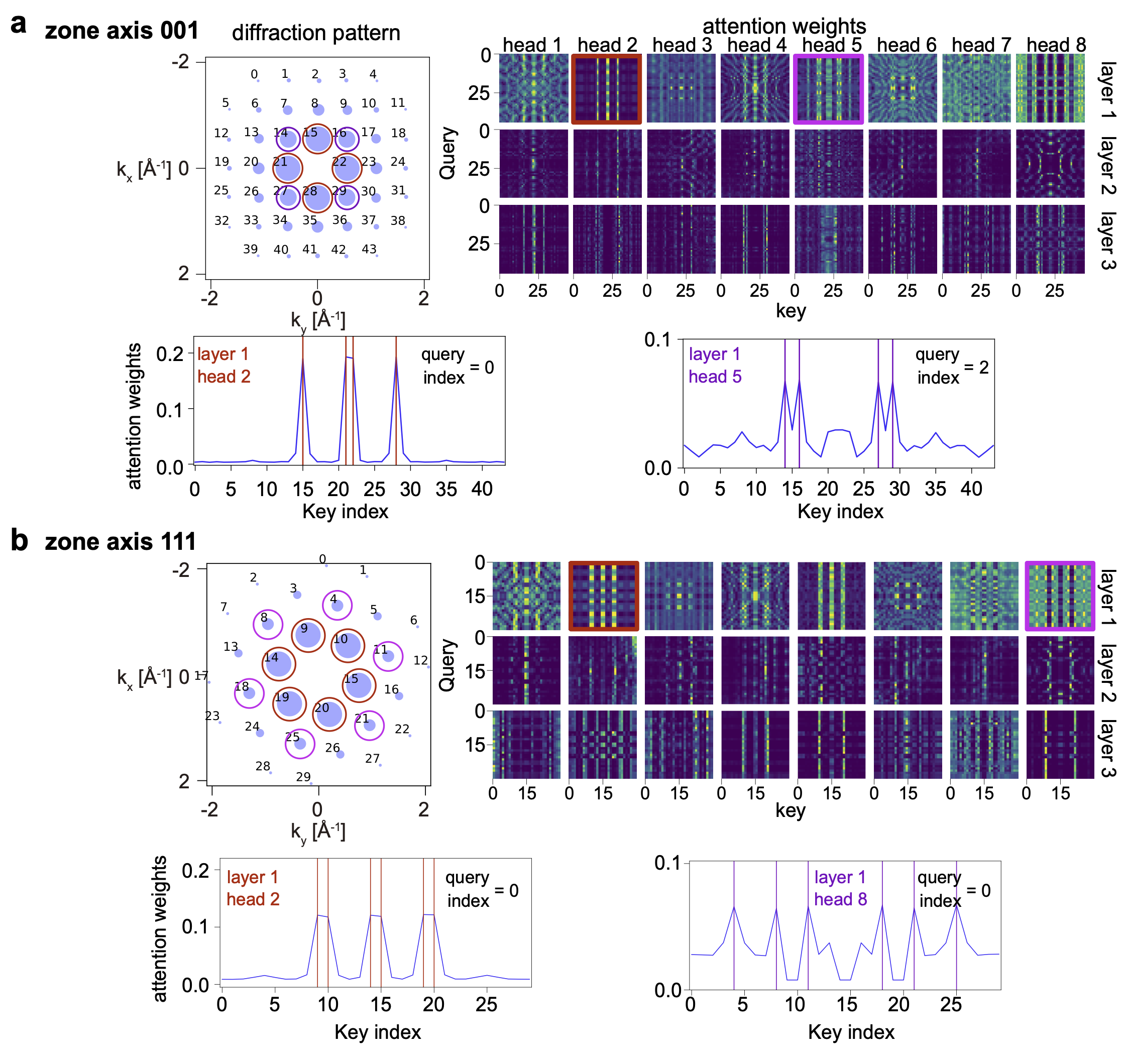}
\vspace*{-3mm}
\caption{\textbf{a} For a diffraction pattern taken along the [001] zone axis (top left), attention weights from all heads and all encoder layers are shown (top right). Two heads from the lowest encoder layer (highlighted in brown and purple) show striped patterns. (bottom) Representative attention-weight profiles along the key indices corresponding to these striped patterns. For each profile, high attention weights are observed at four key indices corresponding to Bragg disks related by four-fold rotational symmetry (top left). \textbf{b} Corresponding plot for a diffraction pattern taken along the [111] zone axis. The two highlighted heads exhibit striped patterns, with higher attention weights at six key indices related to (pseudo-)six-fold rotational symmetry.}
\label{fig:FIG_S18}
\end{center}
\end{figure}

\begin{figure}[htb]
\begin{center}
\includegraphics[width=0.99\textwidth]{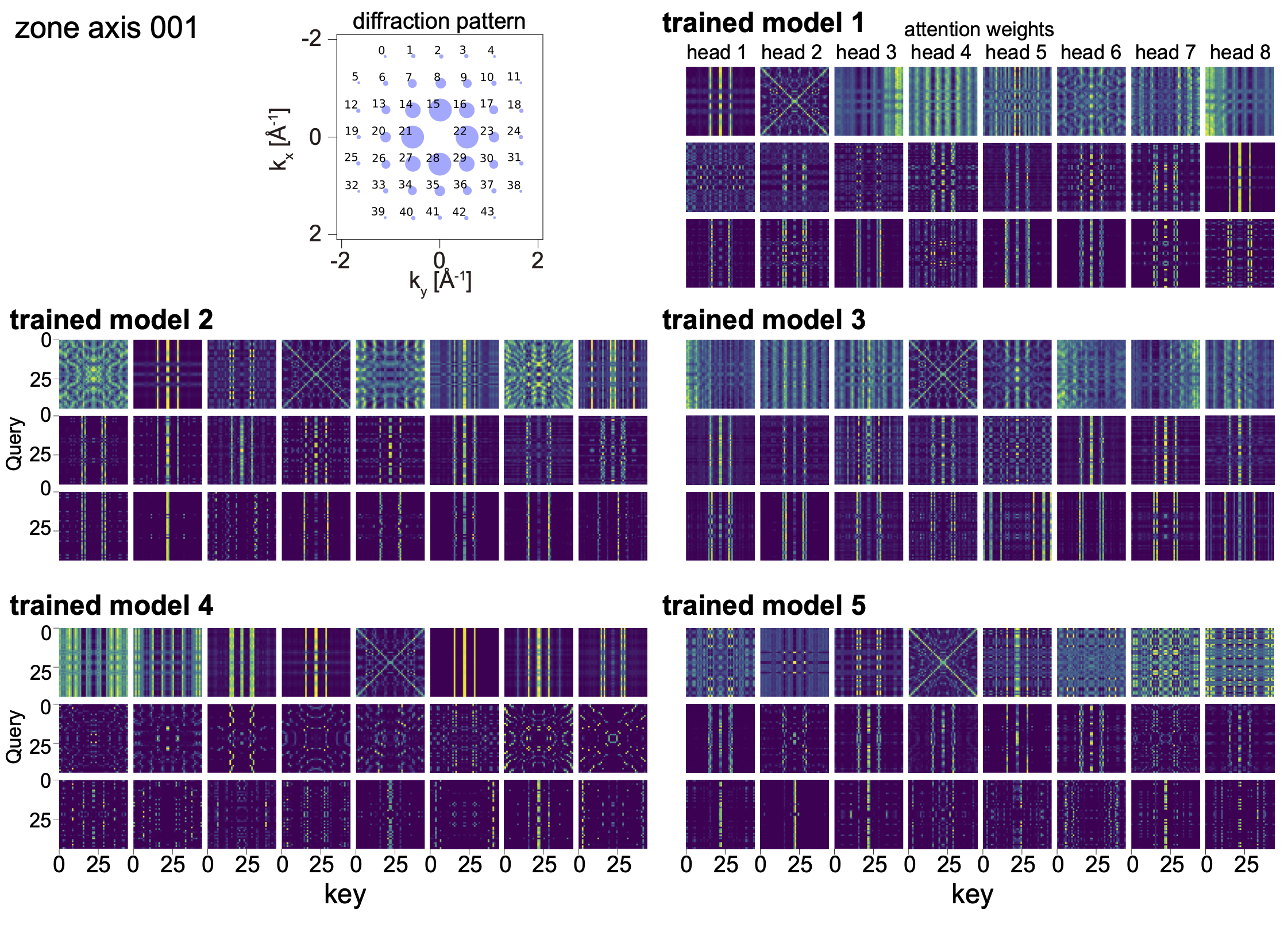}
\vspace*{-3mm}
\caption{For a diffraction pattern taken along the [001] zone-axis direction (top left), the corresponding attention weights from five independently trained models are shown. These models were trained in addition to the model used to generate the attention weights shown in Fig.S13--18.}
\label{fig:FIG_SX19}
\end{center}
\end{figure}

\begin{figure}[htb]
\begin{center}
\includegraphics[width=0.99\textwidth]{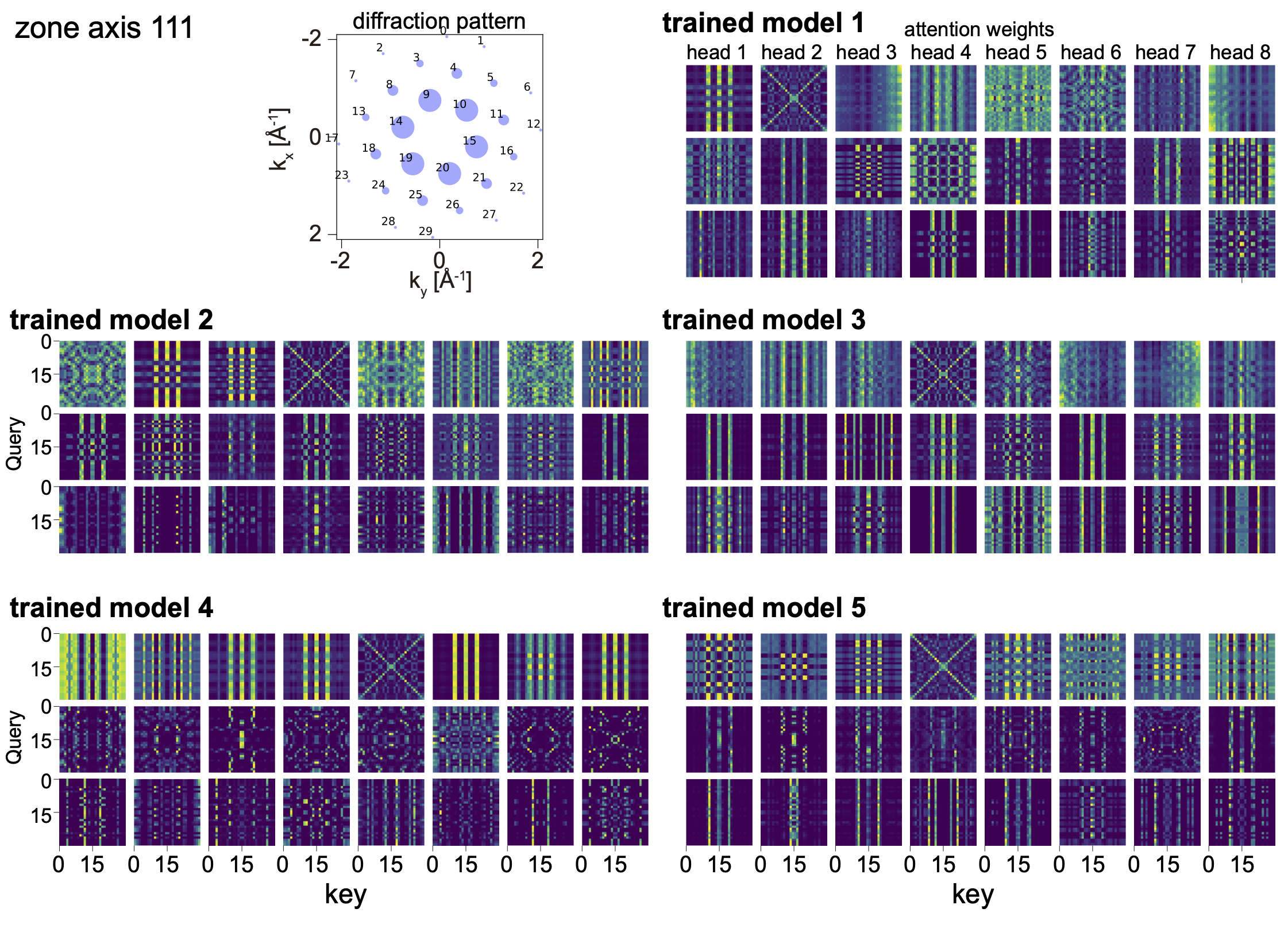}
\vspace*{-3mm}
\caption{For a diffraction pattern taken along the [111] zone-axis direction (top left), the corresponding attention weights from five independently trained models are shown. These models were trained in addition to the model used to generate the attention weights shown in Fig.S13--18.}
\label{fig:FIG_SX20}
\end{center}
\end{figure}

\clearpage

\section*{Supplementary Tables. 1-4}

\begingroup
\singlespacing
\setlength{\tabcolsep}{3pt} 

\begin{table}[H]
\centering
\renewcommand{\arraystretch}{1.4} 
\begin{tabular}{cccccc}
\makebox[3cm][c]{$\begin{bmatrix}
\texttt{ 1 } & \texttt{ 0 } & \texttt{ 0 } \\
\texttt{ 0 } & \texttt{ 1 } & \texttt{ 0 } \\
\texttt{ 0 } & \texttt{ 0 } & \texttt{ 1 }
\end{bmatrix}$}
&
\makebox[3cm][c]{$\begin{bmatrix}
\overline{1} & \texttt{ 0 } & \texttt{ 0 } \\
\texttt{ 0 } & \overline{1} & \texttt{ 0 } \\
\texttt{ 0 } & \texttt{ 0 } & \texttt{ 1 }
\end{bmatrix}$}
&
\makebox[3cm][c]{$\begin{bmatrix}
\overline{1} & \texttt{ 0 } & \texttt{ 0 } \\
\texttt{ 0 } & \texttt{ 1 } & \texttt{ 0 } \\
\texttt{ 0 } & \texttt{ 0 } & \overline{1}
\end{bmatrix}$}
&
\makebox[3cm][c]{$\begin{bmatrix}
\texttt{ 1 } & \texttt{ 0 } & \texttt{ 0 } \\
\texttt{ 0 } & \overline{1} & \texttt{ 0 } \\
\texttt{ 0 } & \texttt{ 0 } & \overline{1}
\end{bmatrix}$} \\[1.0cm]

\makebox[3cm][c]{$\begin{bmatrix}
\texttt{ 0 } & \texttt{ 0 } & \texttt{ 1 } \\
\texttt{ 1 } & \texttt{ 0 } & \texttt{ 0 } \\
\texttt{ 0 } & \texttt{ 1 } & \texttt{ 0 }
\end{bmatrix}$}
&
\makebox[3cm][c]{$\begin{bmatrix}
\texttt{ 0 } & \texttt{ 0 } & \overline{1} \\
\overline{1} & \texttt{ 0 } & \texttt{ 0 } \\
\texttt{ 0 } & \texttt{ 1 } & \texttt{ 0 }
\end{bmatrix}$}
&
\makebox[3cm][c]{$\begin{bmatrix}
\texttt{ 0 } & \texttt{ 0 } & \texttt{ 1 } \\
\overline{1} & \texttt{ 0 } & \texttt{ 0 } \\
\texttt{ 0 } & \overline{1} & \texttt{ 0 }
\end{bmatrix}$}
&
\makebox[3cm][c]{$\begin{bmatrix}
\texttt{ 0 } & \texttt{ 0 } & \overline{1} \\
\texttt{ 1 } & \texttt{ 0 } & \texttt{ 0 } \\
\texttt{ 0 } & \overline{1} & \texttt{ 0 }
\end{bmatrix}$} \\[1.0cm]

\makebox[3cm][c]{$\begin{bmatrix}
\texttt{ 0 } & \texttt{ 1 } & \texttt{ 0 } \\
\texttt{ 0 } & \texttt{ 0 } & \texttt{ 1 } \\
\texttt{ 1 } & \texttt{ 0 } & \texttt{ 0 }
\end{bmatrix}$}
&
\makebox[3cm][c]{$\begin{bmatrix}
\texttt{ 0 } & \overline{1} & \texttt{ 0 } \\
\texttt{ 0 } & \texttt{ 0 } & \texttt{ 1 } \\
\overline{1} & \texttt{ 0 } & \texttt{ 0 }
\end{bmatrix}$}
&
\makebox[3cm][c]{$\begin{bmatrix}
\texttt{ 0 } & \overline{1} & \texttt{ 0 } \\
\texttt{ 0 } & \texttt{ 0 } & \overline{1} \\
\texttt{ 1 } & \texttt{ 0 } & \texttt{ 0 }
\end{bmatrix}$}
&
\makebox[3cm][c]{$\begin{bmatrix}
\texttt{ 0 } & \texttt{ 1 } & \texttt{ 0 } \\
\texttt{ 0 } & \texttt{ 0 } & \overline{1} \\
\overline{1} & \texttt{ 0 } & \texttt{ 0 }
\end{bmatrix}$} \\[1.0cm]

\makebox[3cm][c]{$\begin{bmatrix}
\texttt{ 0 } & \texttt{ 1 } & \texttt{ 0 } \\
\texttt{ 1 } & \texttt{ 0 } & \texttt{ 0 } \\
\texttt{ 0 } & \texttt{ 0 } & \overline{1}
\end{bmatrix}$}
&
\makebox[3cm][c]{$\begin{bmatrix}
\texttt{ 0 } & \overline{1} & \texttt{ 0 } \\
\overline{1} & \texttt{ 0 } & \texttt{ 0 } \\
\texttt{ 0 } & \texttt{ 0 } & \overline{1}
\end{bmatrix}$}
&
\makebox[3cm][c]{$\begin{bmatrix}
\texttt{ 0 } & \overline{1} & \texttt{ 0 } \\
\texttt{ 1 } & \texttt{ 0 } & \texttt{ 0 } \\
\texttt{ 0 } & \texttt{ 0 } & \texttt{ 1 }
\end{bmatrix}$}
&
\makebox[3cm][c]{$\begin{bmatrix}
\texttt{ 0 } & \texttt{ 1 } & \texttt{ 0 } \\
\overline{1} & \texttt{ 0 } & \texttt{ 0 } \\
\texttt{ 0 } & \texttt{ 0 } & \texttt{ 1 }
\end{bmatrix}$} \\[1.0cm]

\makebox[3cm][c]{$\begin{bmatrix}
\texttt{ 0 } & \texttt{ 0 } & \texttt{ 1 } \\
\texttt{ 0 } & \overline{1} & \texttt{ 0 } \\
\texttt{ 1 } & \texttt{ 0 } & \texttt{ 0 }
\end{bmatrix}$}
&
\makebox[3cm][c]{$\begin{bmatrix}
\texttt{ 0 } & \texttt{ 0 } & \overline{1} \\
\texttt{ 0 } & \overline{1} & \texttt{ 0 } \\
\overline{1} & \texttt{ 0 } & \texttt{ 0 }
\end{bmatrix}$}
&
\makebox[3cm][c]{$\begin{bmatrix}
\texttt{ 0 } & \texttt{ 0 } & \texttt{ 1 } \\
\texttt{ 0 } & \texttt{ 1 } & \texttt{ 0 } \\
\overline{1} & \texttt{ 0 } & \texttt{ 0 }
\end{bmatrix}$}
&
\makebox[3cm][c]{$\begin{bmatrix}
\texttt{ 0 } & \texttt{ 0 } & \overline{1} \\
\texttt{ 0 } & \texttt{ 1 } & \texttt{ 0 } \\
\texttt{ 1 } & \texttt{ 0 } & \texttt{ 0 }
\end{bmatrix}$} \\[1.0cm]

\makebox[3cm][c]{$\begin{bmatrix}
\overline{1} & \texttt{ 0 } & \texttt{ 0 } \\
\texttt{ 0 } & \texttt{ 0 } & \texttt{ 1 } \\
\texttt{ 0 } & \texttt{ 1 } & \texttt{ 0 }
\end{bmatrix}$}
&
\makebox[3cm][c]{$\begin{bmatrix}
\overline{1} & \texttt{ 0 } & \texttt{ 0 } \\
\texttt{ 0 } & \texttt{ 0 } & \overline{1} \\
\texttt{ 0 } & \overline{1} & \texttt{ 0 }
\end{bmatrix}$}
&
\makebox[3cm][c]{$\begin{bmatrix}
\texttt{ 1 } & \texttt{ 0 } & \texttt{ 0 } \\
\texttt{ 0 } & \texttt{ 0 } & \overline{1} \\
\texttt{ 0 } & \texttt{ 1 } & \texttt{ 0 }
\end{bmatrix}$}
&
\makebox[3cm][c]{$\begin{bmatrix}
\texttt{ 1 } & \texttt{ 0 } & \texttt{ 0 } \\
\texttt{ 0 } & \texttt{ 0 } & \texttt{ 1 } \\
\texttt{ 0 } & \overline{1} & \texttt{ 0 }
\end{bmatrix}$} \\

\end{tabular}
\caption{Matrices of proper point group symmetry operations for a cubic crystal system.}
\end{table}
\endgroup

\clearpage

\newcommand{\numfont}{\fontsize{6.6pt}{7.5pt}\selectfont}
\begin{table}[h]
\centering
\small
\renewcommand{\arraystretch}{1.25}
\setlength{\tabcolsep}{0.1pt}  
\begin{adjustbox}{width=1.05\textwidth,center}
\begin{tabularx}{\textwidth}{c|*{12}{>{\numfont\Centering\arraybackslash}X|}}
\toprule
\multicolumn{13}{c}{\textbf{Template matching}} \\
\midrule
 & Elapsed (s) & User CPU (s) & Sys CPU (s)
 & Elapsed (s) & User CPU (s) & Sys CPU (s)
 & Elapsed (s) & User CPU (s) & Sys CPU (s)
 & Elapsed (s) & User CPU (s) & Sys CPU (s) \\
\midrule
Run ID
 & \multicolumn{3}{c|}{Scan grid 8$\times$8}
 & \multicolumn{3}{c|}{Scan grid 16$\times$16}
 & \multicolumn{3}{c|}{Scan grid 32$\times$32}
 & \multicolumn{3}{c}{Scan grid 64$\times$64} \\
\midrule
1 & 6.79 & 9.99 & 1.16 & 10.12 & 12.78 & 1.70 & 24.07 & 24.41 & 4.01 & 81.17 & 71.55 & 13.93 \\
2 &  3.46 & 8.06 & 0.53 & 7.20 & 11.24 & 1.08 & 20.93 & 22.65 & 3.39 & 76.65 & 69.18 & 12.50 \\
3 &  3.48 & 8.10 & 0.50 & 7.12 & 11.10 & 1.14 & 21.00 & 22.86 & 3.26 & 77.42 & 70.06 & 12.45 \\
4 &  3.28 & 7.90 & 0.51 & 7.04 & 11.07 & 1.09 & 21.83 & 23.35 & 3.59 & 77.02 & 69.98 & 12.15 \\
5 &  3.51 & 8.13 & 0.49 & 6.87 & 10.90 & 1.10 & 21.17 & 22.75 & 3.54 & 76.77 & 69.57 & 12.29 \\
6 &  3.32 & 7.97 & 0.47 & 6.79 & 10.87 & 1.05 & 21.02 & 22.84 & 3.30 & 77.30 & 70.11 & 12.26 \\
7 &  3.55 & 8.16 & 0.52 & 6.99 & 11.05 & 1.07 & 21.15 & 22.72 & 3.54 & 77.62 & 69.26 & 13.45 \\
8 &  3.44 & 8.06 & 0.50 & 7.02 & 11.07 & 1.07 & 21.09 & 22.68 & 3.55 & 77.92 & 69.49 & 13.51 \\
9 &  3.43 & 8.06 & 0.51 & 6.67 & 10.63 & 1.10 & 21.21 & 22.78 & 3.54 & 78.19 & 69.94 & 13.35 \\
10 &  3.56 & 8.19 & 0.50 & 6.95 & 10.92 & 1.16 & 21.01 & 22.54 & 3.51 & 77.12 & 69.16 & 13.07 \\
\midrule
Run ID
 & \multicolumn{3}{c|}{Scan grid 128$\times$128}
 & \multicolumn{3}{c|}{Scan grid 256$\times$256}
 & \multicolumn{3}{c|}{Scan grid 512$\times$512}
 & \multicolumn{3}{c}{Scan grid 1024$\times$1024} \\
\midrule
1 & 305.81 & 260.43 & 49.62 & 1317.27 & 975.79 & 345.22 & 5343.06 & 3839.67 & 1505.60 & 21389.10 & 15352.72 & 6032.42  \\
2 &  303.95 & 256.33 & 52.63 & 1297.47 & 963.32 & 338.73 & 5358.78 & 3851.54 & 1510.16 & 21397.11 & 15386.89 & 6006.68  \\
3 &  303.17 & 255.20 & 52.99 & 1318.73 & 964.96 & 358.34 & 5365.51 & 3853.89 & 1514.56 & 21356.43 & 15345.34 & 6007.48  \\
4 &  305.05 & 257.36 & 52.69 & 1339.22 & 971.55 & 372.21 & 5355.00 & 3845.62 & 1512.27 & 21490.12 & 15429.45 & 6056.94  \\
5 &  301.82 & 258.45 & 48.38 & 1348.33 & 976.14 & 376.72 & 5188.96 & 3818.00 & 1373.96 & 21396.44 & 15321.16 & 6071.40  \\
6 &  300.04 & 256.18 & 48.87 & 1341.95 & 971.05 & 375.43 & 5302.79 & 3818.83 & 1486.90 & 21255.87 & 15268.35 & 5983.86  \\
7 &  301.51 & 257.67 & 48.84 & 1336.07 & 975.86 & 364.71 & 5185.13 & 3838.90 & 1349.18 & 21458.17 & 15391.02 & 6063.44  \\
8 &  303.47 & 259.94 & 48.54 & 1349.13 & 981.54 & 372.15 & 5173.86 & 3824.48 & 1352.37 & 21478.29 & 15422.18 & 6052.23  \\
9 &  300.17 & 257.06 & 48.12 & 1346.21 & 975.11 & 375.59 & 5338.85 & 3852.09 & 1489.70 & 21413.02 & 15357.69 & 6051.56  \\
10 &  303.06 & 256.74 & 51.33 & 1334.94 & 968.60 & 370.89 & 5285.99 & 3801.52 & 1487.45 & 21364.30 & 15344.63 & 6015.96 \\
\bottomrule
\end{tabularx}
\end{adjustbox}
\caption{Computation time of py4DSTEM template matching methods for multiple scan grids. For each scan grid, elapsed wall time (left), user CPU time (middle), and system CPU time (right) were measured over 10 independent serial runs.}
\label{tab:template-matching}
\end{table}
\clearpage

\begin{table}[h]
\centering
\small
\renewcommand{\arraystretch}{1.25}
\setlength{\tabcolsep}{0.1pt}  
\begin{adjustbox}{width=1.05\textwidth,center}
\begin{tabularx}{\textwidth}{c|*{12}{>{\numfont\Centering\arraybackslash}X|}}
\toprule
\multicolumn{13}{c}{\textbf{Model prediction (CPU-only)}} \\
\midrule
 & Elapsed (s) & User CPU (s) & Sys CPU (s)
 & Elapsed (s) & User CPU (s) & Sys CPU (s)
 & Elapsed (s) & User CPU (s) & Sys CPU (s)
 & Elapsed (s) & User CPU (s) & Sys CPU (s) \\
\midrule
Run ID
 & \multicolumn{3}{c|}{Scan grid 8$\times$8}
 & \multicolumn{3}{c|}{Scan grid 16$\times$16}
 & \multicolumn{3}{c|}{Scan grid 32$\times$32}
 & \multicolumn{3}{c}{Scan grid 64$\times$64} \\
\midrule
1 & 3.76 & 9.22 & 0.59 & 4.06 & 14.36 & 1.37 & 5.38 & 29.58 & 5.30 & 9.04 & 85.15 & 19.57 \\
2 & 2.49 & 8.73 & 0.43 & 2.79 & 13.89 & 1.27 & 3.79 & 29.27 & 5.16 & 7.77 & 85.15 & 19.59 \\
3 & 2.51 & 8.87 & 0.41 & 2.76 & 13.39 & 1.27 & 3.79 & 29.41 & 5.23 & 7.75 & 84.41 & 19.88 \\
4 & 2.52 & 8.75 & 0.44 & 2.74 & 13.41 & 1.22 & 3.78 & 29.34 & 5.25 & 7.79 & 84.79 & 19.90 \\
5 & 2.53 & 8.84 & 0.45 & 2.86 & 14.19 & 1.29 & 3.82 & 29.88 & 5.14 & 7.84 & 85.72 & 19.67 \\
6 & 2.50 & 8.76 & 0.40 & 2.81 & 13.96 & 1.30 & 3.79 & 29.45 & 5.35 & 7.73 & 84.39 & 19.77 \\
7 & 2.52 & 8.80 & 0.44 & 2.78 & 13.59 & 1.31 & 3.83 & 29.69 & 5.23 & 7.91 & 85.68 & 19.97 \\
8 & 2.52 & 8.80 & 0.40 & 2.84 & 14.14 & 1.32 & 3.84 & 29.80 & 5.19 & 7.86 & 85.57 & 19.64 \\
9 & 2.51 & 8.88 & 0.43 & 2.82 & 14.13 & 1.29 & 3.84 & 29.49 & 5.28 & 7.79 & 84.99 & 19.70 \\
10 & 2.53 & 8.73 & 0.45 & 2.79 & 13.65 & 1.29 & 3.87 & 29.85 & 5.21 & 7.84 & 85.43 & 20.09 \\
\midrule
Run ID
 & \multicolumn{3}{c|}{Scan grid 128$\times$128}
 & \multicolumn{3}{c|}{Scan grid 256$\times$256}
 & \multicolumn{3}{c|}{Scan grid 512$\times$512}
 & \multicolumn{3}{c}{Scan grid 1024$\times$1024} \\
\midrule
1 & 24.70 & 314.86 & 77.62 & 87.43 & 1243.02 & 312.48 & 348.17 & 5053.26 & 1264.08 & 1419.70 & 21311.64 & 4116.99 \\
2 & 23.51 & 315.36 & 78.09 & 88.45 & 1272.52 & 313.18 & 351.43 & 5091.30 & 1273.70 & 1433.07 & 21314.13 & 4119.84 \\
3 & 23.50 & 315.44 & 78.39 & 88.40 & 1267.49 & 315.31 & 353.06 & 5121.76 & 1275.87 & 1444.28 & 21360.55 & 4142.01 \\
4 & 23.93 & 321.95 & 78.64 & 88.91 & 1269.09 & 317.26 & 355.31 & 5123.00 & 1277.15 & 1450.37 & 21328.49 & 4154.08 \\
5 & 24.01 & 320.83 & 78.72 & 88.99 & 1273.75 & 315.73 & 354.34 & 5121.66 & 1283.75 & 1451.78 & 21340.23 & 4163.19 \\
6 & 24.07 & 322.56 & 79.08 & 89.10 & 1280.42 & 315.99 & 355.93 & 5130.98 & 1284.33 & 1457.35 & 21251.77 & 4174.52 \\
7 & 24.07 & 324.56 & 79.47 & 89.31 & 1279.61 & 315.48 & 356.90 & 5132.72 & 1288.95 & 1455.47 & 21241.80 & 4168.74 \\
8 & 23.85 & 319.66 & 78.33 & 88.79 & 1270.34 & 315.84 & 357.26 & 5131.44 & 1285.00 & 1456.74 & 21256.70 & 4177.99 \\
9 & 24.06 & 322.40 & 80.35 & 88.90 & 1276.67 & 316.70 & 356.78 & 5139.53 & 1282.16 & 1464.16 & 21311.26 & 4188.38 \\
10 & 24.18 & 322.90 & 80.11 & 89.12 & 1280.63 & 316.13 & 355.89 & 5125.88 & 1293.37 & 1461.31 & 21298.83 & 4185.24 \\
\bottomrule
\end{tabularx}
\end{adjustbox}
\caption{Computation time of CPU-only model prediction for multiple scan grids. For each scan grid, elapsed wall time (left), user CPU time (middle), and system CPU time (right) were measured over 10 independent serial runs.}
\label{tab:template-matching}
\end{table}
\clearpage

\begin{table}[h]
\centering
\small
\renewcommand{\arraystretch}{1.25}
\setlength{\tabcolsep}{0.1pt}  
\begin{adjustbox}{width=1.05\textwidth,center}
\begin{tabularx}{\textwidth}{c|*{12}{>{\numfont\Centering\arraybackslash}X|}}
\toprule
\multicolumn{13}{c}{\textbf{Model prediction (GPU-accelerated)}} \\
\midrule
 & Elapsed (s) & User CPU (s) & Sys CPU (s)
 & Elapsed (s) & User CPU (s) & Sys CPU (s)
 & Elapsed (s) & User CPU (s) & Sys CPU (s)
 & Elapsed (s) & User CPU (s) & Sys CPU (s) \\
\midrule
Run ID
 & \multicolumn{3}{c|}{Scan grid 8$\times$8}
 & \multicolumn{3}{c|}{Scan grid 16$\times$16}
 & \multicolumn{3}{c|}{Scan grid 32$\times$32}
 & \multicolumn{3}{c}{Scan grid 64$\times$64} \\
\midrule
1 & 4.02 & 7.76 & 0.68 & 4.03 & 7.93 & 0.74 & 4.17 & 8.04 & 0.74 & 4.80 & 8.88 & 0.74 \\
2 & 2.72 & 7.61 & 0.53 & 2.74 & 7.81 & 0.53 & 2.91 & 8.14 & 0.49 & 3.44 & 8.74 & 0.50 \\
3 & 2.71 & 7.56 & 0.53 & 2.71 & 7.79 & 0.50 & 2.83 & 7.97 & 0.50 & 3.52 & 8.79 & 0.56 \\
4 & 2.68 & 7.61 & 0.49 & 2.70 & 7.84 & 0.49 & 2.89 & 8.07 & 0.51 & 3.54 & 8.81 & 0.54 \\
5 & 2.73 & 7.67 & 0.49 & 2.72 & 7.79 & 0.52 & 2.93 & 8.14 & 0.50 & 3.46 & 8.76 & 0.55 \\
6 & 2.71 & 7.68 & 0.50 & 2.74 & 7.94 & 0.52 & 2.89 & 8.05 & 0.51 & 3.43 & 8.78 & 0.51 \\
7 & 2.68 & 7.63 & 0.51 & 2.71 & 7.78 & 0.51 & 2.90 & 8.00 & 0.51 & 3.47 & 8.82 & 0.53 \\
8 & 2.72 & 7.56 & 0.53 & 2.72 & 7.80 & 0.53 & 2.90 & 8.08 & 0.49 & 3.47 & 8.76 & 0.53 \\
9 & 2.72 & 7.68 & 0.51 & 2.75 & 7.87 & 0.52 & 2.88 & 8.07 & 0.51 & 3.52 & 8.78 & 0.59 \\
10 & 2.74 & 7.82 & 0.47 & 2.71 & 7.81 & 0.51 & 2.90 & 8.11 & 0.50 & 3.43 & 8.73 & 0.51 \\
\midrule
Run ID
 & \multicolumn{3}{c|}{Scan grid 128$\times$128}
 & \multicolumn{3}{c|}{Scan grid 256$\times$256}
 & \multicolumn{3}{c|}{Scan grid 512$\times$512}
 & \multicolumn{3}{c}{Scan grid 1024$\times$1024} \\
\midrule
1 & 7.13 & 12.07 & 0.75 & 16.38 & 24.54 & 0.87 & 52.88 & 74.03 & 1.40 & 202.25 & 275.33 & 4.77 \\
2 & 5.89 & 12.09 & 0.55 & 14.95 & 24.33 & 0.67 & 52.02 & 74.53 & 1.21 & 199.28 & 273.77 & 3.92 \\
3 & 5.74 & 11.89 & 0.52 & 14.89 & 24.27 & 0.70 & 51.71 & 74.22 & 1.18 & 199.72 & 274.47 & 3.65 \\
4 & 5.78 & 11.88 & 0.58 & 14.93 & 24.29 & 0.67 & 51.91 & 74.53 & 1.21 & 197.80 & 272.38 & 3.60 \\
5 & 5.76 & 11.88 & 0.54 & 15.03 & 24.44 & 0.67 & 51.92 & 74.35 & 1.22 & 199.48 & 273.97 & 4.19 \\
6 & 5.78 & 11.97 & 0.54 & 15.05 & 24.38 & 0.67 & 52.33 & 74.82 & 1.18 & 199.51 & 274.11 & 4.30 \\
7 & 5.79 & 11.91 & 0.57 & 15.03 & 24.47 & 0.67 & 51.85 & 74.35 & 1.20 & 200.68 & 275.34 & 3.87 \\
8 & 5.87 & 12.07 & 0.56 & 15.11 & 24.51 & 0.68 & 51.66 & 74.11 & 1.25 & 198.18 & 272.85 & 3.70 \\
9 & 5.79 & 11.87 & 0.58 & 14.95 & 24.40 & 0.67 & 52.11 & 74.54 & 1.24 & 199.37 & 273.97 & 4.31 \\
10 & 5.77 & 11.88 & 0.58 & 15.06 & 24.53 & 0.69 & 51.80 & 74.18 & 1.30 & 199.50 & 274.05 & 3.87 \\
\bottomrule
\end{tabularx}
\end{adjustbox}
\caption{Computation time of GPU-accelerated model prediction for multiple scan grids. For each scan grid, elapsed wall time (left), user CPU time (middle), and system CPU time (right) were measured over 10 independent serial runs.}
\label{tab:template-matching}
\end{table}
\clearpage

\section*{Legends for Mvoies S1-S2}

\subsection*{Movie S1}
Paired Bragg disks related by Friedel symmetry for a diffraction pattern taken along the [001] zone axis (top), and the corresponding attention weights assigned by the transformer model across all eight attention heads and three encoder layers (bottom). Attention is evaluated separately, with each Bragg disk treated as a query. Red labels adjacent to the Bragg disks denote their Miller indices. In the lowest encoder layer, Bragg disks related by Friedel symmetry show similar attention patterns across key indices.

\clearpage

\bibliography{main}
\bibliographystyle{naturemag}